\documentclass[fleqn,usenatbib,onecolumn]{mnras}
\usepackage{natbib}
\usepackage{newtxtext,newtxmath}
\usepackage{enumitem}
\usepackage{float}
\restylefloat{table}

\usepackage[T1]{fontenc}
\usepackage{ae,aecompl}
\usepackage{setspace}
\renewcommand\vec{\mathbf}
\newcommand*\Bell{\ensuremath{\boldsymbol\ell}}
\makeatletter
\newcommand\footnoteref[1]{\protected@xdef\@thefnmark{\ref{#1}}\@footnotemark}
\makeatother

\usepackage{graphicx}	
\usepackage{amsmath}	
\usepackage{amssymb}	
\usepackage[dvipsnames]{xcolor}

\newcommand{\mr}[1]{\mathrm{#1}}
\def\Msun{\ensuremath{M_\odot}}

\title[Fast weak lensing bispectrum covariance]{Tomographic weak lensing bispectrum: a thorough analysis towards the next generation of galaxy surveys}
 
\author[Matteo Rizzato et al.]{
Matteo Rizzato,$^{1,2}$\thanks{E-mail: matteo.rizzato@iap.fr}
Karim Benabed,$^{1,2}$
Francis Bernardeau,$^{1,2,3}$
Fabien Lacasa$^{4}$
\\
$^{1}$Sorbonne Universit\'e, CNRS, UMR 7095, Institut d'Astrophysique de Paris, 98 bis bd Arago, 75014 Paris, France\\
$^{2}$Sorbonne Universit\'es, Institut Lagrange de Paris (ILP), 98 bis bd Arago, 75014 Paris, France\\
$^{3}$Institut de Physique Th\'eorique, Universit\'e Paris-Saclay, CEA, CNRS, UMR3681, 91191 Gif-sur-Yvette, France, France\\
$^{4}$D\'{e}partement de Physique Th\'{e}orique and Center for Astroparticle Physics, Universit\'{e} de Gen\`{e}ve, 24 quai Ernest Ansermet, \\\hspace{0.3cm}-1211 Geneva, Switzerland
}

\date{Accepted 2019 October 6. Received  2019 October 2; in original form 2019 February 22}

\pubyear{2018}

\begin{document}
\label{firstpage}
\pagerange{\pageref{firstpage}--\pageref{lastpage}}
\maketitle

\begin{abstract}
We address key points for an efficient implementation of likelihood codes for modern weak lensing large-scale structure surveys. Specifically, we focus on the joint  weak lensing convergence power spectrum-bispectrum probe and we tackle the numerical challenges required by a realistic analysis. Under the assumption of (multivariate) Gaussian likelihoods, we have developed a high performance code that allows highly parallelised prediction of the binned tomographic observables and of their joint non-Gaussian covariance matrix accounting for terms up to the 6-point correlation function and super-sample effects. This performance allows us to qualitatively address several interesting scientific questions.
We find that the bispectrum provides an improvement in terms of signal-to-noise ratio (S/N) of about 10\% on top of the power spectrum, making it a non-negligible source of information for future surveys.
Furthermore, we are capable to test the impact of theoretical uncertainties in the halo model used to build our observables; with presently allowed variations we conclude that the impact is negligible on the S/N.
Finally, we consider data compression possibilities to optimise future analyses of the weak lensing bispectrum. We find that, ignoring systematics, 5 equipopulated redshift bins are enough to recover the information content of a Euclid-like survey, with negligible improvement when increasing to 10 bins. We also explore principal component analysis and dependence on the triangle shapes as ways to reduce the numerical complexity of the problem.
\end{abstract}

\begin{keywords}
cosmology: theory - cosmology: large-scale structure of Universe - gravitational lensing: weak - methods: analytical - methods: statistical
\end{keywords}



\section{Introduction}
\label{Intro}
In the coming decades, large-scale structure (LSS) surveys will probe the late time Universe with an unprecedented precision, allowing us to constrain dark energy models. Euclid, the Dark Energy Survey (DES) and the Large Synoptic Survey Telescope (LSST) are well known examples of observational programs aiming at this scientific goal \citep{2011arXiv1110.3193L,2005astro.ph.10346T,2009arXiv0912.0201L}. One of their key probes will be the detection of the weak lensing signal from distant sources. Unlike galaxy clustering measurements, weak lensing probes the total amount of matter in the Universe \citep{2001PhR...340..291B,2015RPPh...78h6901K} giving us unbiased information on its distribution and evolution over time.
In preparation for future missions, we need to understand the performance of this probe in terms of cosmological parameter error forecasts, crucially for deviations away from the $\Lambda$CDM scenario.

A standard way to extract cosmological information from the matter field, is through the computation of its 2-point correlation function and associated errors. So far, weak lensing analyses have employed this basic approach, the field of interest being the projection of the matter field along the line-of-sight. At low redshift, the non-linear evolution of the LSS of the Universe skews the distribution of the matter field and the 2-point correlation function is not sufficient for a complete description of the statistical properties of the field. In order to recover the cosmological information not accessible via standard approaches, different techniques have been advocated \citep{2010PhRvD..81d3519K,2013MNRAS.434.2961C,2010PhRvD..82b3002B,2015PhRvD..92h3531P,2019arXiv190902615B}. In this work, we exploit the higher order correlation functions of the weak lensing field to catch part of the missing cosmological information. We rely on the assumption that their joint statistical distribution is a multivariate Gaussian.

The constraining power of this approach can be addressed either via Fisher forecast \citep{tegmark1997measuring}, DALI forecast~\citep{2014MNRAS.441.1831S,2015MNRAS.453..893S} or Markov Chain Monte Carlo (MCMC) analyses \citep{2001CQGra..18.2677C,2002PhRvD..66j3511L,2005MNRAS.356..925D,2013ascl.soft03003A}.
The core part of these approaches lies in the computation (and inversion) of covariance matrices to quantify the correlations between different binned modes of the field. As for weak lensing, modern galaxy surveys will be capable to detect the position of the sources in different tomographic redshift bins probing the time evolution of the matter field. The complexity of our covariance matrices is enhanced by the correlations between observables sourced by galaxies in different bins: with the final set-up considered in our computations, we compute covariance matrices with $\gtrsim 10^{4}\times 10^{4}$ elements. Future data analyses based on the joint weak lensing power spectrum-bispectrum probe requires the manipulation of these matrices. Optimisation via the reduction of their dimensionality is an extremely important path to explore~\citep{2017MNRAS.472.4244H}. Bearing this idea in mind, we present possible summary statistics to be applied to both the power spectrum and the bispectrum and we explore possible paths to further simplify the complexity of future forecasts.
In order to make these analyses feasible (at least for a small cluster), an original algorithm had to be developed: 
the code will shortly be made available publicly, together with its documentation.

The weak lensing signal is sensitive to very small scales. For a Euclid-like survey, \cite{2011MNRAS.416.1717K} proved that the power spectrum has to be accurately known to 1\% down to $k \approx 50 h\ \mathrm{Mpc}^{-1}$ to saturate the dark energy figure of merit. This study is in agreement with the previous one from \cite{2009arXiv0905.0501D}. Similarly, \cite{2005APh....23..369H} argued that we typically need the power spectrum within a few percent accuracy at the $k \approx 10 h\ \mathrm{Mpc}^{-1}$ in order for the uncertainty on the power spectrum itself not to degrade cosmological constraints in DES- and LSST-like survey. In \cite{2011MNRAS.418..536E,2012JCAP...04..034H} a similar conclusion is obtained.  At these scales linear theory for the evolution of the matter perturbations can not be trusted to build our observables.
Perturbative approaches (e.g. perturbation theory up to 2 loops) starts deviating by more than $1\%$ at scales $k \geq 0.1 h\ \mathrm{Mpc}^{-1}$ at $z = 0$ \citep{PhysRevD.86.103528}. In the above regime cosmological N-body simulations are usually employed to study the non-linear gravitational evolution and eventually used to tune phenomenological models or fitting formulae for the power spectrum. For example, \cite{1996MNRAS.280L..19P} provided a fitting formula for the power spectrum based on a scaling ansatz presented in \cite{1991ApJ...374L...1H}. Later, \cite{2003MNRAS.341.1311S} proposed a new model of the power spectrum, the so-called halofit model, which is based on a the well known halo model of structure formation (e.g. \cite{2000ApJ...543..503M} ; \cite{2000MNRAS.318..203S}; \cite{2001ApJ...554...56C}). \cite{2015MNRAS.454.1958M} presented an optimised variant of the halo model, designed to produce accurate matter power spectra well into the non-linear regime for a wide range of cosmological models, including baryonic feedback. The halo model, along with its variants, has nowadays become a standard in the computation of covariance matrices within galaxy survey pipelines \citep{2017MNRAS.470.2100K,2017MNRAS.465.1454H}. All the approaches mentioned above are meant to provide a good fit for the power spectrum alone. However, they give a poor description of the true underlying physics \citep{2018arXiv181102976R}. In particular, they cannot be used to compute higher order statistics, which are key ingredients in our analysis. Here, we rather build our observables in the standard halo model framework. Recently,  \cite{2013MNRAS.429..344K} proved that the halo model has a precision of $20 \%$ up to scales $k \approx 10 h\ \mathrm{Mpc}^{-1}$ at $z = 0$ in terms of weak lensing convergence power spectrum and bispectrum and of the joint covariance. Though not sufficient for actual data analysis, this precision is
sufficient for our purpose of a S/N evaluation. While we perform several approximations to optimise the computational resources at our disposal, the mentioned model accuracy is actually the real bottleneck of our implementation.
Even if promising, the halo model is not the final theory for dark matter clustering: its precision might start degrading before reaching the desired scales, assumptions have to be used to build the observables and different effects can be added or neglected according to their importance for the forecast.
Thanks to the high performance of the tool developed within the present project, we test the robustness of our forecast against one of the main uncertainty of the model. We underline that the main goal of this work is the validation and the test of the algorithm developed for the computation and compression of the joint weak lensing convergence power spectrum-bispectrum covariance matrix, being a challenging numerical task by itself. We leave sophisticated parameter forecasts (which anyway require the evaluation of the covariance) for future works. Instead we use here a signal-to-noise ratio (S/N) analysis as suggested by \cite{1997ApJ...480...22T}, which has become a standard tool in the literature to gauge the information content of weak lensing observables \citep{2005MNRAS.360L..82R,2009ApJ...701..945S,2009MNRAS.395.2065T,2013MNRAS.429..344K}.

This paper is organised as follows. In Sec.~\ref{sec:WLintro}, we introduce our notation for the theory of weak lensing. In Sec.~\ref{sec:CorrCov} we provide the expressions for the observables used in this work (power spectrum and bispectrum of weak lensing convergence) and for their covariance matrix. We also list further parameters related to the modelling of the latter. The expressions provided in this section are meant to be generic: no use of any specific physical model is required. In Sec.~\ref{sec:halomodelmain} we give a quick review of the halo model and we build the three dimensional spectra in this frame: at this point, all the assumptions and the analytical expressions required for our implementation are in place. Sec.~\ref{sec:analysis}  and Sec.~\ref{sec:PCA} contain the core part of this paper. After introducing the joint S/N analysis for the power spectrum and the bispectrum, we derive our results and analyse the possibility for future data compression via the principal component analysis (PCA) of our covariance matrices. Finally, in Sec.~\ref{sec:Conclusions} we summarise our main findings and the approximations underlying our analyses. 

The tools we developed are extremely flexible and can be applied to different experimental setup. In this work, we derive our results according to the specificities of the ESA Euclid mission~\citep{2011arXiv1110.3193L}, listed in Appendix~\ref{App:EuclidSpec}.
We build our observables in flat-sky exploiting the Limber approximation when moving to the statistics of the fields projected along the line-of-sight. The good performance of this approximation has been well investigated at the level of power spectra for weak lensing in \cite{2017MNRAS.472.2126K}: we assume that this result is reliable also at the level of higher order correlation functions given that we are interested in the information content at small scales. Verifying these approximations, however, is an interesting path to follow in future works.
We assume a 6-parameters spatially flat $\Lambda$CDM with ${\Omega_{\mathrm{m}}} = 0.238$, ${\Omega_{\mathrm{b}}} = 0.042$, ${\Omega_{\Lambda}} = 0.762$, $h=0.732$, $\sigma_8=0.76$, $n_{\mathrm{s}}=0.958$  and we work in units $c=1$. Lengths and masses are respectively measured in $\mathrm{Mpc}/h$ and $\mathrm{M}_{\odot}/h$.
\section{Weak Lensing}
\label{sec:WLintro}

The images of far sources are distorted due to the underlying matter distribution.
The matter fractional density field $\delta$ locally perturbs the background metric through the Weyl scalar potential. Given this perturbed metric, we can solve the geodesic equation for a photon emitted in a given angular direction $\boldsymbol{\theta}$ at a specific time, and evaluate its angular deviation from an unperturbed trajectory at every moment. This deviation is directly related to the first derivative of the post-newtonian gravitational potentials. The magnification of the sources in a redshift bin $b_i = \left[z_i, z_{i+1}\right]$ induced by the cosmological weak lensing in a given angular direction can be estimated via the tomographic convergence field $\kappa_{(i)}\left(\boldsymbol{\theta}\right)$, which is in particular related to the Laplacian of the potential. The full calculation proves that we can express this field as the line-of-sight integration of the matter density contrast, properly convoluted with a geometrical kernel which accounts for the position of the sources~\citep{2001PhR...340..291B,2015RPPh...78h6901K}. Precisely, for sources selected in a redshift bin $b_i$, the convergence reads
\begin{equation}
\label{kappa}
\kappa_{(i)}\left(\boldsymbol{\theta}\right) = \int_0^{\chi(z_{i+1})} \mathrm{d}\chi \mathcal{W}_{(i)} \left( \chi \right) \delta\left[\boldsymbol{\theta}\chi, \chi\right] 
\end{equation}
where $\chi\left(z\right)$ is the comoving distance from the observer and $H\left( z \right)$ is the time dependent Hubble factor. The function $\mathcal{W}_{(i)}\left( \chi \right)$ is the aforementioned lensing kernel associated to the projected distribution of the sources placed within the $i^{\text{th}}$ redshift bin. We express this last quantity as the convolution of a cosmology (only)-dependent function $\mathcal{S}\left(z_{\mathrm{s}},z\right)$ and a function $\mathcal{F}\left(z_{\mathrm{s}},i\right)$ which is dependent on the properties of the survey
\begin{equation}
\label{KernelLensing1}
\mathcal{W}_{(i)}\left( \chi\left(z\right) \right) = \int_{z}^{\infty} \mathrm{d}z_{\mathrm{s}}\ \mathcal{F}\left(z_{\mathrm{s}},i\right) \mathcal{S}\left(z_{\mathrm{s}},z\right).
\end{equation}
In Eq.~\eqref{KernelLensing1}, $z$ is the redshift that enters the line-of-sight integration in Eq.~\eqref{kappa} and $z_{\mathrm{s}}$ is the one of the sources whose contribution to the signal is considered. The function $\mathcal{S}\left(z_{\mathrm{s}},z\right)$ is also known as lensing efficiency
\begin{equation}
\label{LensEff}
\mathcal{S}\left(z_{\mathrm{s}},z\right) = \frac{3}{2}\Omega_{\mathrm{m},0}H_0^2\left(1+z\right)\chi\left(z\right)\frac{\chi\left(z_{\mathrm{s}}\right)-\chi\left(z\right)}{\chi\left(z_{\mathrm{s}}\right)}  
\end{equation}
while, $\mathcal{F}\left(z_{\mathrm{s}},i\right)$ is simply the expected projected number density of the sources in the $i^{\text{th}}$ bin $n_i\left(z_{\mathrm{s}}\right)$ and it is vanishing outside the considered bin. In a more general approach, this last function should also account for the photometric errors in the detection of the source positions. Given the complexity of the computation we perform, we omit this contribution assuming we are capable to measure the realistic position of the sources without any errors. 

\section{Correlation functions and covariance estimation}
\label{sec:CorrCov}
\subsection{Higher order correlation functions for weak lensing}
Convergence maps exhibit large-scale correlations that reflect those of the matter field.
One of the main results of this work consists in the analysis of the level of correlation between power spectra and bispectra for the tomographic weak lensing convergence over different scales and spatial configurations. In Sec.~\ref{3.2} and in Appendix~\ref{App:covariances}, we show that we need the correlation functions of the convergence field up to the 6-point one. Indeed, the non-linear gravitational evolution couples different modes of the matter field leading to non-vanishing higher order correlation functions. 
In Fourier space, we define the connected part of the $n^{\text{th}}$ order correlation function of the tomographic convergence field \eqref{kappa}
\begin{equation}
\label{CorrFunct}
\langle\kappa^{(i_1)}_{\boldsymbol\ell_1} \dots \kappa^{(i_n)}_{\boldsymbol\ell_n}\rangle_c \equiv \left(2\pi\right)^2 P_{i_1,\dots,i_n}\left(\boldsymbol\ell_1, \dots,\boldsymbol\ell_n\right) \delta_{\mathrm{D}}\left(\boldsymbol\ell_1 + \dots + \boldsymbol\ell_n\right)
\end{equation}
where $\kappa^{(i)}_{\boldsymbol{\ell}} = \int d\boldsymbol{\theta}\ \kappa_{(i)}\left(\boldsymbol{\theta}\right) e^{-i\boldsymbol{\ell}\cdot\boldsymbol{\theta}}$ is the Fourier transform of the convergence field \eqref{kappa} and $\delta_{\mathrm{D}}$ is the Dirac delta. We call the quantity $P_{i_1,\dots,i_n}\left(\boldsymbol\ell_1, \dots,\boldsymbol\ell_n\right)$ polyspectrum of order $n$. In Eq.~\eqref{CorrFunct} we described the field on a flat sky, i.e. we approximate the full spherical harmonics decomposition of the real field with a simple two-dimensional Fourier transform. Such approximation is valid at percent level for $\ell > 100$ \citep{2017MNRAS.472.2126K}. By replacing the convergence field definition into Eq.~\eqref{CorrFunct}, we can derive the expression for the flat-sky polyspectra. In general, the redshift integration (line-of-sight integration) appearing in Eq.~\eqref{kappa} would naturally translate into a complex $n$-dimensional one. To simplify this calculation, we make use of the Limber approximation \citep{PhysRevD.78.123506}: we assume that the three-dimensional matter polyspectra have a weak dependence on the momenta component corresponding to the line-of-sight direction. 
Consequently, the projection collapses into a simple one-dimension redshift integration and we can relate the angular multipoles $\boldsymbol{\ell}$ to the three-dimensional momenta $\mathbf{k}$ via the well known Limber relation $\mathbf{k}\left(\Bell,z\right)\approx \{\Bell / \chi\left(z\right),0\}$. Finally, the general $n$-order tomographic convergence polyspectrum $P_{i_1,\dots,i_n} \left(\boldsymbol{\ell}_1, \dots,\boldsymbol{\ell}_n\right)$ relates to the same order matter one $P\left(\mathbf{k}_1, \dots,\mathbf{k}_n\right)$ via 
\begin{multline}
P^{(n)}_{i_1\dots i_n} \left(\boldsymbol{\ell}_1, \dots,\boldsymbol{\ell}_n\right) = \int _0^{\infty} \mathrm{d}\chi\ \chi^{2-2n}\ \left[\prod_{\hat{i}=i_1}^{i_n} \mathcal{W}_{(\hat{i})}\left( \chi \right)\right]\  P\left(\mathbf{k}\left(\Bell_1,\chi\right), \dots,\mathbf{k}\left(\Bell_n,\chi\right)\right)   \\
\equiv \int _0^{\infty} \mathrm{d}\chi\ \mathcal{T}\left(i_1,\dots,i_n;\chi\right)  P\left(\mathbf{k}\left(\Bell_1,\chi\right), \dots,\mathbf{k}\left(\Bell_n,\chi\right)\right).
\label{nProjection}
\end{multline}
For consistency with the literature, we call the 2-, the 3- and the 4-order polyspectrum respectively power spectrum, bispectrum and trispectrum
\begin{align}
P_{ij}\left(\ell\right) &\equiv P^{(2)}_{ij} \left(\Bell\right),\label{RenamePS}\\
B_{ijk}\left(\ell_1,\ell_2,\ell_3\right) &\equiv P^{(3)}_{ijk} \left(\Bell_1,\Bell_2,\Bell_3\right),\label{RenameBS}\\
T_{ijkl}\left(\Bell_1,\Bell_2,\Bell_3,\Bell_4\right) &\equiv P^{(4)}_{ijkl} \left(\Bell_1,\Bell_2,\Bell_3,\Bell_4\right).
\end{align}
We underline that the assumption of an isotropic and homogeneous Universe (Cosmological Principle) allows us to reduce the actual dependencies of the polyspectra. The power spectrum is expressed as function of the module $\ell$ of the momentum $\Bell$ (we define $\ell \equiv |\Bell|$) and the bispectrum has a dependence on just 3 degrees of freedom, e.g. the edges of the associated triangular configuration \citep{2017JCAP...02..032M}
The power spectrum measurements for an actual survey are also affected by intrinsic shape noise due to the finite number of sources and the intrinsic variability of galaxy shapes. Assuming that the orientation of intrinsic galaxy shapes is random and the shapes of different galaxies are uncorrelated, this component is Gaussian\footnote{A more realistic case would have a binomial shot noise, leading to a non-vanishing bispectrum component. We leave this consideration to future works.} and we account for this effect in the following way
\begin{equation}
\label{ShotNoise}
P_{ij}^{\text{s.}}\left(\ell\right) \equiv P_{ij}\left(\ell\right) + \frac{\sigma_{\epsilon}^2}{\bar{n}_{(i)}} \, \delta^{\mathrm{K}}_{ij} .
\end{equation}
 In Eq.~\eqref{ShotNoise},  $\bar{n}_{(i)}$ is the expected projected number of sources per unit of solid angle within the $i^{\mathrm{th}}$ redshift bin and $\delta^{\mathrm{K}}_{ij}$ is the Kronecker delta for the two indices $i,j$.
In the following numerical calculations, we use the value $\sigma_{\epsilon} = 0.3$ which is representative of the expected Euclid sample \citep{2013LRR....16....6A} and we ignore weak lensing systematics like intrinsic alignments \citep{2010MNRAS.402.2127S,2013MNRAS.436..819J,2015PhR...558....1T,2015SSRv..193...67K,2017arXiv170809247B,2018JCAP...07..030S}, photometric redshift errors \citep{2006ApJ...636...21M}, blending \citep{2011A&A...528A..51H} and Point Spread Function \citep{2008JCAP...01..003J,2013PASJ...65..104H,2013MNRAS.428.2695C}.

\subsection{Covariance matrix for the observables}
\label{3.2}
\subsubsection{Structure of the covariance matrix}
In this work, we want to address the cosmological information content of the joint tomographic convergence power spectrum-bispectrum probe. We accomplish this task via a signal-to-noise ratio analysis (described in Sec.~\ref{sec:analysis}) for which a data vector of estimators and the corresponding covariance matrix are required. For our computation, we chose the binned estimators $\hat{P}^{\mathrm{W}}_{ij}(\ell^{\mathrm{b}})$ \eqref{PSWest} and  $\hat{B}^{\mathrm{W}}_{ijk}(\ell_1^{\mathrm{b}},\ell_2^{\mathrm{b}},\ell_3^{\mathrm{b}})$ \eqref{BSWset} for the true underlying tomographic power spectrum \eqref{RenamePS} and bispectrum \eqref{RenameBS} respectively. We bin the magnitude $\ell$ of the angular momenta $\Bell$ in equally-spaced bins in $\log\ell$. At the level of notation, we define the bin $\ell^{\mathrm{b}}$ as the symmetric interval of width $\Delta\ell^{\mathrm{b}}$ around the central value $\ell$. Therefore, a multipole $\Bell'$ belongs to the bin $\ell^{\mathrm{b}}$ when its magnitude $\ell'\in \left[\ell-\Delta\ell^{\mathrm{b}}/2,\ell+\Delta\ell^{\mathrm{b}}/2\right]$. We refer to Appendix~\ref{App:EuclidSpec} for a detailed description of the chosen binning.
Moving to the covariance of the estimators, we split it in the following way \citep{2001ApJ...554...56C,2007NJPh....9..446T,2009MNRAS.395.2065T,2009ApJ...701..945S,2013MNRAS.429..344K,2013arXiv1306.4684K}
\begin{align}
\text{Cov}\left[\hat{P}^{\mathrm{W}}_{ij}\left(\ell^{\mathrm{b}}\right) ,\hat{P}^{\mathrm{W}}_{i'j'}\left(\ell^{'\mathrm{b}}\right) \right] &= \text{Cov}\left[\dots \right]_{\text{Gauss}} + \text{Cov}\left[\dots \right]_{\text{NGins}} + \text{Cov}\left[\dots\right]_{\text{NGssc}},\label{CovPPsynt}\\
\text{Cov}\left[\hat{B}^{\mathrm{W}}_{ijk}\left(\ell_1^{\mathrm{b}},\ell_2^{\mathrm{b}},\ell_3^{\mathrm{b}}\right),\hat{B}^{\mathrm{W}}_{i'j'k'}\left(\ell^{'\mathrm{b}}_1,\ell^{'\mathrm{b}}_2,\ell^{'\mathrm{b}}_3\right) \right] &=\text{Cov}\left[\dots\right]_{\text{Gauss}} + \text{Cov}\left[\dots\right]_{\text{NGins}} +
\text{Cov}\left[\dots \right]_{\text{NGssc}}, \label{CovBBsynt}\\
\text{Cov}\left[\hat{P}^{\mathrm{W}}_{ij}\left(\ell^{\mathrm{b}}\right),\hat{B}^{\mathrm{W}}_{i'j'k'}\left(\ell^{\mathrm{b}}_1,\ell^{\mathrm{b}}_2,\ell^{\mathrm{b}}_3\right) \right] &=
\text{Cov}\left[\dots\right]_{\text{NGins}} +
\text{Cov}\left[\dots\right]_{\text{NGssc}}. \label{CovPBsynt}
\end{align}
In Eqs.~\eqref{CovPPsynt} and \eqref{CovPBsynt}, we label with the subscript Gauss the covariance terms containing only 2-point statistics, which are non-vanishing only for correlations within the same $\Bell-$bin. The other covariance terms arise due to the non-Gaussian statistics of the convergence field and correlate modes in different $\Bell-$bins and the different probes, i.e. power spectrum and bispectrum. We distinguish two classes of terms, respectively labelled via the subscripts NGins and NGssc. The former is sourced by correlations between observed intra-survey modes, while the latter is sourced by correlations between observed modes and background super-survey modes and it is known in the literature as super-sample covariance (SSC) \citep{10.1111/j.1365-2966.2006.10709.x,2009MNRAS.395.2065T,2013PhRvD..87l3504T,2018JCAP...06..015B,2018PhRvD..97d3532C,2019JCAP...03..008B}. For notation purpose, in Appendix~\ref{App:covariances} we will further split the NGins-like terms in different components according to the order of the correlations sourcing them. Also, we will generically dub as NG the total non-Gaussian covariance, regardless of the type of correlations involved. In the following section we will briefly review the main logical and mathematical steps needed for deriving the covariance matrix for the power spectrum \eqref{CovPPsynt}, in all its components. As not to break the flow of the paper, we defer to Appendix~\ref{App:covariances} a similar analysis for the bispectrum and the power spectrum-bispectrum cross-covariance.

\subsubsection{Power spectrum covariance matrix: Gauss and NGins terms}
We give here few details about the computation of the power spectrum covariance while deferring the discussion about the bispectrum and power spectrum-bispectrum cross-covariance to Appendix~\ref{App:covariances}. We start by defining the binned estimator for tomographic power spectrum
\begin{equation}
\label{PSest}
    \hat{P}_{ij}(\ell^{\mathrm{b}}) \equiv \frac{1}{\Omega_{\text{sky}}\  N\left(\ell^{\mathrm{b}}\right)}\sum_{\Bell'}\kappa_{\Bell'}^{(i)}\kappa_{-\Bell'}^{(j)}\Delta^{(2)}_{\ell^{\mathrm{b}}}\left(\Bell'\right).
\end{equation}
The sum in Eq.~\eqref{PSest} runs over discrete modes which are integer multiples of the fundamental frequency of our survey $\ell_{\mathrm{f}} \equiv 2\pi/\Theta_{\text{sky}}$, $\Theta_{\text{sky}}$ being the survey footprint angular size. In particular, the finite real-space domain of our observation implies that we cannot measure angular modes $\ell\le\ell_{\mathrm{f}}$ and also $\ell_{\mathrm{f}}$ provides a minimum resolution for the module of the multipoles we can access
\begin{equation}
     \Bell \sim \Bell_{n_x}^{n_y} = \ell_{\mathrm{f}}\cdot \{n_x,n_y\},\quad n_x,\ n_y\ \in \ \mathbb{N}^{\times}. 
\end{equation} 
The estimator \eqref{PSest} differs from the one introduced in Eqs.~\eqref{CovPPsynt} and \eqref{CovPBsynt}. In particular the estimator  \eqref{PSest} does not account for the impact of the survey mask function, whose effect is considered when introducing the estimator $\hat{P}^{\mathrm{W}}_{ij}(\ell^{\mathrm{b}})$ \eqref{PSWest}. The link between the two will be clearer in the following. We choose this presentation for didactic purposes.
The selection function $\Delta^{(2)}_{\ell^{\mathrm{b}}}\left(\Bell'\right) = 1$ when the magnitude $\ell'$ of the mode $\Bell'$ falls into the required bin $\ell^{\mathrm{b}}$. Also, the quantity $N(\ell^{\mathrm{b}})\approx 2\ell\Delta\ell^{\mathrm{b}} f_{\text{sky}}$ (in the limit $\ell\gg\ell_f$) gives the number of  vector pairs $\hat{\Bell},-\hat{\Bell}$ whose magnitude $\hat{\ell}$ is within the bin $\ell^{\mathrm{b}}$, each pair being discriminated by a deviation in the module of the vectors of a unit of the survey fundamental mode $\ell_{\mathrm{f}}$ \citep{2008A&A...477...43J,2009A&A...508.1193J,2013MNRAS.429..344K}.
In Eq.~\eqref{PSest} we have introduced the survey angular coverage $\Omega_{\text{sky}}$, which is related to the linear angular size of the survey $\Theta_{\text{sky}}$ via
\begin{equation}
    \Omega_{\text{sky}} = 2\pi\left(1-\cos\Theta_{\text{sky}}\right).
\end{equation}
Often, it is useful to refer to the angular coverage in terms of the fraction $f_{\text{sky}}$ of the sky which is observed by the survey. In our case, it is defined as 
\begin{equation}
f_{\text{sky}} \equiv \frac{\Omega_{\text{sky}}}{4\pi}.
\end{equation} 
The estimator \eqref{PSest} is proved unbiased \citep{2007NJPh....9..446T}. Also, assuming that the true power spectrum varies slowly within the bin width, it can be approximated by the power spectrum \eqref{RenamePS} itself as calculated at the central value of the corresponding bin, i.e.
\begin{equation}
    \langle\hat{P}_{ij}(\ell^{\mathrm{b}})\rangle \approx P_{ij}(\ell).
\end{equation}
We can derive the covariance of the estimator~\eqref{PSest} by applying its standard definition \citep{2001ApJ...554...56C,2007NJPh....9..446T,2009MNRAS.395.2065T,2009ApJ...701..945S,2013MNRAS.429..344K} 
\begin{multline}
\label{CovPP1}
\text{Cov}\left[\hat{P}_{ij}(\ell^{\mathrm{b}}) ,\hat{P}_{i'j'}(\ell^{'\mathrm{b}}) \right] \equiv 
\langle \hat{P}_{ij}(\ell^{\mathrm{b}})\hat{P}_{i'j'}(\ell^{'\mathrm{b}}) \rangle -\langle \hat{P}_{ij}(\ell^{\mathrm{b}})\rangle\langle \hat{P}_{i'j'}(\ell^{'\mathrm{b}})\rangle = \\
 \frac{1}{\Omega_{\text{sky}}\  N\left(\ell^{\mathrm{b}}\right)} \frac{1}{\Omega_{\text{sky}}\  N\left(\ell^{'\mathrm{b}}\right)}\sum_{\bar{\Bell},\bar{\Bell}'}\langle\kappa_{\bar{\Bell}}^{(i)}\kappa_{-\bar{\Bell}}^{(j)}\kappa_{\bar{\Bell}'}^{(i')}\kappa_{-\bar{\Bell}'}^{(j')}\rangle\Delta^{(2)}_{\ell^{\mathrm{b}}}\left(\bar{\Bell}\right)\Delta^{(2)}_{\ell^{'\mathrm{b}}}(\bar{\Bell}') - P_{ij}(\ell) P_{i'j'}(\ell^{'}).
\end{multline}
In particular, within the covariance, we account for the shot noise term introduced in Eq.~\eqref{ShotNoise}.
The Gaussian component can be easily derived via the Wick's theorem by decomposing the 4-point correlator in Eq.~\eqref{CovPP1} into products of power spectra. A detailed calculation leads to
\begin{equation}
\text{Cov}\left[\hat{P}_{ij}\left(\ell^{\mathrm{b}}\right) ,\hat{P}_{i'j'}\left(\ell^{'\mathrm{b}}\right) \right]_{\text{Gauss}} = \frac{\delta^{\mathrm{K}}_{\ell\ell'}}{N\left(\ell^{\mathrm{b}}\right)}\left[ P_{ii'}^{\text{s.}}\left(\ell\right) P_{jj'}^{\text{s.}}\left(\ell\right) +P_{ij'}^{\text{s.}}\left(\ell\right) P_{ji'}^{\text{s.}}\left(\ell\right) \right].\label{CovPPG}
\end{equation}
The connected component leads instead to the NGins term in Eq.~\eqref{CovPPsynt}
\begin{equation}
\text{Cov}\left[\hat{P}_{ij}(\ell^{\mathrm{b}}) ,\hat{P}_{i'j'}(\ell^{'\mathrm{b}})\right]_{\text{NGins}} = \frac{1}{N\left(\ell^{\mathrm{b}}\right)N\left(\ell^{'\mathrm{b}}\right)\Omega_{\text{sky}}}
   \sum_{\bar{\Bell},\bar{\Bell}'}
   \left[T_{iji'j'}\left(\bar{\Bell},-\bar{\Bell},\bar{\Bell}',\bar{\Bell}'\right)
   \right]
   \Delta^{(2)}_{\ell^{\mathrm{b}}}\left(\bar{\Bell}\right)\Delta^{(2)}_{\ell^{'\mathrm{b}}}(\bar{\Bell}')\approx \frac{1}{\Omega_{\mathrm{sky}}}T_{iji'j'}\big(\ell,-\ell,\ell',-\ell'\big).\label{CovPPNG}
\end{equation}
In Eq.~\eqref{CovPPNG} the exact covariance evaluation would require an average of the trispectrum over the two bins $\ell^{\mathrm{b}}, {\ell'}^{\mathrm{b}}$. However, as we will explain in Sec.~\ref{sec:halomodel}, we work in a regime for which the 1-halo term is a good approximation to the trispectrum. This component does not depend on the angles between the wavevectors. On top of that, we assume that the trispectrum does not vary significantly within the bins, dropping the average over the modules as well. 

\subsubsection{Power spectrum covariance matrix: NGssc term}
The super-sample covariance term, i.e. the second non-Gaussian component in Eq.~\eqref{CovPPsynt}, is rooted in the intrinsic nature of our observations. As a matter of fact, the presence (as it is the case in all observations) of a finite survey mask function $W(\boldsymbol{\theta})$ ($\tilde{W}(\Bell)$ in Fourier space) of angular size $\Theta_{\text{sky}}$ induces correlations with modes which are of the order of the fundamental length of its domain in Fourier space, i.e. the fundamental frequency $\ell_{\mathrm{f}}$. 
This component cannot be derived from the estimator~\eqref{PSest} since it does not account for couplings with modes $\ell\leq\ell_f$. We can introduce a more sophisticated one accounting for the impact of the window function $W\left(\boldsymbol{\theta}\right)$ on the observed field
\begin{equation}
\label{PSWest}
\hat{P}_{ij}^{W}(\ell^{\mathrm{b}}) \equiv \frac{1}{\Omega_{\text{sky}}\  N\left(\ell^{\mathrm{b}}\right)}\sum_{\Bell'}{\kappa^W}_{\Bell'}^{(i)}{\kappa^W}_{-\Bell'}^{(j)}\Delta^{(2)}_{\ell^{\mathrm{b}}}\left(\Bell'\right)
\end{equation}
where $\kappa^W$ is the Fourier transform of the observed convergence field: the convolution of the underlying convergence field $\kappa$ with the window function of the survey $\tilde{W}$
\begin{equation}
\label{smoothedF}
    \kappa^{\mathrm{W}}_{(i)}\left(\boldsymbol{\theta}\right) = \mathrm{W}\left(\boldsymbol{\theta}\right)\kappa_{(i)}\left(\boldsymbol{\theta}\right),\qquad
    \kappa^{\mathrm{W}(i)}_{\Bell} = \int\ \frac{\mathrm{d}^2\Bell'}{\left( 2\pi\right)^2}\tilde{\mathrm{W}}\left(\Bell'\right)\kappa^{(i)}_{\Bell-\Bell'}.
\end{equation}
At the observed scales $\ell\gg\ell_{\mathrm{f}}$, a detailed computation proves that the estimator $\hat{P}_{ij}^{W}(\ell^{\mathrm{b}}) $ is unbiased \citep{2013PhRvD..87l3504T,2018JCAP...06..015B} so that
\begin{equation}
    \langle\hat{P}_{ij}^{W}(\ell^{\mathrm{b}}) \rangle \approx P_{ij}\left(\ell\right),
\end{equation}
where once again we approximated the value of the estimator with the power spectrum itself on the central value of the corresponding bin (under the assumption of negligible variations within the bin).
At the level of covariance we have \citep{2013PhRvD..87l3504T,2018JCAP...06..015B,2019JCAP...03..008B}
\begin{align}
    &\text{Cov}\left[\hat{P}^W_{ij}(\ell^{\mathrm{b}}) ,\hat{P}^W_{i'j'}(\ell^{'\mathrm{b}}) \right]_{\text{Gauss}} \approx \text{Cov}\left[\hat{P}_{ij}(\ell) ,\hat{P}_{i'j'}(\ell') \right]_{\text{Gauss}},\\
    &\text{Cov}\left[\hat{P}^W_{ij}\left(\ell^{\mathrm{b}}\right) ,\hat{P}^W_{i'j'}(\ell^{'\mathrm{b}}) \right]_{\text{NG}} \approx \frac{1}{N\left(\ell^{\mathrm{b}}\right)N(\ell^{'\mathrm{b}})\Omega_{\text{sky}}}
   \sum_{\bar{\Bell},\bar{\Bell}'}\int\frac{\mathrm{d}^2\Bell^{''}}{\left( 2\pi\right)^2}|W(\Bell^{''})|^2T_{iji'j'}\big(\Bell,-\Bell+\Bell^{''},\Bell',-\Bell'-\Bell^{''}\big)\Delta^{(2)}_{\ell^{\mathrm{b}}}\left(\bar{\Bell}\right)\Delta^{(2)}_{\ell^{'\mathrm{b}}}(\bar{\Bell}')\label{here}
\end{align}
always in the limit $\ell,\ell'\gg\ell_f$. In Eq.~\eqref{here}, the subscript NG refer to the general non-Gaussian contribution to the covariance regardless of the type of correlations.
Since we are working in the Limber approximation, the trispectrum appearing in Eq.~\eqref{here} can be derived via the projection of the three-dimensional matter one
\begin{equation}
    T_{iji'j'}\big(\Bell,-\Bell+\Bell^{''},\Bell',-\Bell'-\Bell^{''}\big) = \int _0^{\infty} \mathrm{d}\chi\ \chi^{-6}\ \mathcal{T}\left(i,j,i',j';\chi\right) T\left[\vec{k}(\Bell,z),-\vec{k}(\Bell,z)+\vec{k}(\Bell^{''},z),\vec{k}(\Bell',z),-\vec{k}(\Bell',z)-\vec{k}(\Bell^{''},z)\right].
\end{equation}
The three-dimensional momenta are evaluated through the Limber relation introduced around Eq.~\eqref{nProjection}.
At every redshift, the matter trispectrum can be evaluated via the consistency relations introduced by \cite{2013PhRvD..87l3504T}
\begin{equation}
\label{TakadaAnsatz}
T\left(\vec{k}_1,-\vec{k}_1+\vec{p},\vec{k}_2,-\vec{k}_2-\vec{p}\right) \approx
T\left(\vec{k}_1,-\vec{k}_1,\vec{k}_2,-\vec{k}_2\right) + \frac{\partial P(\vec{k}_1|\delta_{\mathrm{b}})}{\partial\delta_{\mathrm{b}}}\frac{\partial P(\vec{k}_2|\delta_{\mathrm{b}})}{\partial\delta_{\mathrm{b}}}\ P^{\text{lin}}\left(p\right).
\end{equation}
The quantities $\partial P/\partial\delta_{\mathrm{b}}$ are the power spectrum responses to a change in the background matter density induced by a long background mode $\delta_{\mathrm{b}}$ 
\begin{equation}
\label{BackgroundMode}
    \delta_{\mathrm{b}}\equiv \frac{1}{V}\int_{\mathrm{v}} d^3\vec{x}\ \delta_{\mathrm{D}}\left(\vec{x}\right).
\end{equation}
$V$ being the volume accessible by the survey. Matter responses, such as those required in Eq.~\eqref{TakadaAnsatz}, can be either measured from simulations or computed within a given theoretical framework via the so called separate Universe ansatz \citep{2005ApJ...634..728S,2011JCAP...10..031B,2012PhRvD..85j3523S,2014PhRvD..89h3519L,2015MNRAS.448L..11W,2016JCAP...09..007B}. We employ the second approach and we briefly go through their derivation as from the halo model in Sec.~\ref{subsubResp}.   
After replacing Eq.~\eqref{TakadaAnsatz} within Eq.~\eqref{here}, the first term in Eq.~\eqref{TakadaAnsatz} easily leads to the expression for the intra-survey covariance NGins in Eq.~\eqref{CovPPNG}. The second term instead involves couplings with super-sample modes $p\ \chi\leq\ell_{\mathrm{f}}$, commonly assumed to be in the linear regime 
\begin{align}
  \text{Cov}\left[\hat{P}^W_{ij}(\ell^{\mathrm{b}}) ,\hat{P}^W_{i'j'}(\ell^{'\mathrm{b}}) \right]_{\text{NGssc}} &\approx \frac{1}{N\left(\ell^{\mathrm{b}}\right)N\left(\ell^{'\mathrm{b}}\right)\Omega_{\text{sky}}}
   \int _0^{\infty} \mathrm{d}\chi\ \chi^{-6}\ \mathcal{T}\left(i,j,i',j';\chi\right)\times\nonumber\\
   &\hspace{2cm}\sum_{\bar{\Bell},\bar{\Bell}'}\frac{\partial P(\vec{k}(\bar{\Bell}/\chi)|\delta_{\mathrm{b}})}{\partial\delta_{\mathrm{b}}}\frac{\partial P(\vec{k}(\bar{\Bell}',\chi)|\delta_{\mathrm{b}})}{\partial\delta_{\mathrm{b}}}\Delta^{(2)}_{\ell^{\mathrm{b}}}\left(\bar{\Bell}\right)\Delta^{(2)}_{\ell^{'\mathrm{b}}}(\bar{\Bell}')\int\frac{\mathrm{d}^2\Bell^{''}}{\left( 2\pi\right)^2}|\tilde{W}(\Bell^{''})|^2 P^{\text{lin}}\left(\vec{k}(\Bell^{''},\chi)\right) \nonumber\\
   & \approx \frac{1}{\Omega_{\text{sky}}}
   \int _0^{\infty} \mathrm{d}\chi\ \chi^{-6}\ \mathcal{T}\left(i,j,i',j';\chi\right)\frac{\partial P(\vec{k}(\Bell,\chi)|\delta_{\mathrm{b}})}{\partial\delta_{\mathrm{b}}}\frac{\partial P(k(\Bell',\chi)|\delta_{\mathrm{b}})}{\partial\delta_{\mathrm{b}}}\sigma_W^2\left(\chi\right).\label{CovPP22}
\end{align}
We assumed slowly varying responses within the bins. The quantity $\sigma_W^2\left(\chi\right)$ is the time-dependent variance of the linearly-evolved matter field within the observed disk-like volume at comoving distance $\chi$ from the observer
\begin{equation}
    \sigma_W^2\left(\chi\right)\equiv \int\frac{\mathrm{d}^2\Bell}{\left( 2\pi\right)^2}|\tilde{W}(\Bell)|^2 P^{\text{lin}}\left(k(\Bell,\chi)\right).
\end{equation}
We will give a detailed expression of this term in Appendix~\ref{App:covariances}.

\section{The halo model}
\label{sec:halomodelmain}
\subsection{Halo model performance}
\label{sec:halomodel}
The evaluation of Eqs.~\eqref{CovPPsynt}-\eqref{CovPBsynt} requires the computation of the polyspectra \eqref{CorrFunct} up to the 6\textsuperscript{th} order one plus the responses of the observables to the background modes in the case of the super-sample terms. The halo model \citep{2002PhR...372....1C} provides an effective and physically motivated ansatz to compute the different polyspectra we need. As anticipated in Sec.~\ref{Intro}, this theoretical framework has already been used in the context of weak lensing power spectrum and bispectrum error estimation 
and we expect it to be accurate at $20\%$ up to $\ell\approx 7000$ (see \cite{2013MNRAS.429..344K} for a test without source tomography) with a minor dependence on the details of the implementation (see for example \cite{2005ApJ...632...29F} for a discussion at the level of 2- and 3-point correlation function). Even if the above stated precision may not be enough for the final target of future galaxy surveys, however the halo model represents the best available choice in the literature to model the matter clustering up to the 6-point correlation function from an analytical prospective. Therefore, we consider the above performance satisfactory and in particular good enough for the S/N analysis performed in Sec.~\ref{sec:analysis}. Furthermore, the halo model, along with its variants, has nowadays become a standard in the computation of covariance matrices within galaxy survey pipelines \citep{2017MNRAS.470.2100K,2017MNRAS.465.1454H}. This description of matter clustering relies on the fact, supported by numerical simulations, that we can model the statistical properties of the matter field via halos of dark matter of different masses, redshifts and positions. For the practical halo model implementation, we used the Sheth and Tormen mass function (\cite{1999MNRAS.308..119S}), the NFW halo profile (\cite{1996ApJ...462..563N}), and the concentration-mass relation $c_{\mathrm{v}}-m$ from~\cite{2001MNRAS.321..559B}. In the following we investigate the impact of the uncertainties of the parameters in the $c_{\mathrm{v}}-m$ relation on our covariances. Our computational framework is flexible enough to allow for the estimation of the impact of other assumptions of the model, such the parametrization of the mass function and  the halo profile. This will be left for subsequent works.
In the halo model the $n$-point correlation function is described as the sum of terms accounting for all the possible distributions of the points within the halos: from the 1-halo term where all the points lie within a single halo, dominant at the smallest scales, up to the $n$-halo term where every point lies in a different halo, dominant at the largest scales. For the data vector, we use all the $n$-halo terms, both for the power spectrum (1- and 2-halo) and bispectrum (1-, 2- and 3-halo). We show their analytical expressions in Sec.~\ref{DataVecHM}. We restrict the polyspectra beyond the second order one (power spectrum) within the covariance to the 1-halo term only, the expressions being provided in Sec.~\ref{CovHM} instead. In the simpler case of the bispectrum only covariance, the 2- and 3-halo terms contribute mainly by increasing the correlations between squeezed triangular configurations (i.e. $B(\ell_i,\ell_j,\ell_k)$ with $\ell_k\sim\ell_j\gg\ell_i$), and change the overall signal-to-noise ratio by a few percent. Those squeezed configurations have a low information content. Beside, we verified in the joint covariance that they are strongly correlated with power spectrum modes, and thus does not contribute significantly in our final signal-to-noise figure. For this reason, we simplify our computations and assume that the 1-halo approximation is sufficient for these analyses. 

\subsection{The halo model: matter polyspectra for the data vector}
\label{DataVecHM}
\subsubsection{Halo bias}
In the halo model framework the matter polyspectra are computed as a sum over all the possible point allocations within one or more halos. While the value of the matter field is related to the halo properties via the halo density profile, the statistical properties of the contrast matter density field do not reflect directly those of the number density of halos: matter halos are indeed biased tracers of the underlying matter density field. If we define the halo density contrast for halos of mass $m$ as
\citep{1996MNRAS.282..347M,1997MNRAS.284..189M}
\begin{equation}
\label{DeltaH}
    \delta_{\mathrm{h}}\left(\mathbf{x},t,m\right) \equiv \frac{n_{\mathrm{h}}\left(\mathbf{x},t,m\right)}{\bar{n}_{\mathrm{h}}\left(t,m\right)} - 1
\end{equation}
where $\bar{n}_{\mathrm{h}}\left(t,m\right)$ is the mean comoving number density for halos of mass $m$ and $n_{\mathrm{h}}\left(\vec{x},t,m\right)$ is the actual one at position $\vec{x}$, then $\delta_{\mathrm{h}}\left(\vec{x},t,m\right)$ is not equal to $\delta\left(\vec{x},t\right)$ and it can be expressed as an expansion in powers of the latter (see \cite{2018PhR...733....1D} for a thorough review on the subject). In order to describe the power spectrum and the bispectrum at leading order in perturbation theory (PT) (see \cite{2002PhR...367....1B} for a thorough review), we need terms up the second power in the field $\delta$. In particular, two local biases $b_1(m,t), b_2(m,t)$ \citep{1993ApJ...413..447F} and a non-local bias term $b_{s_2}(m,t)$  \citep{2012PhRvD..85h3509C,2012PhRvD..86h3540B} are required
\begin{equation}
    \label{ExpansionBiasF}
    \delta_{\mathrm{h}}\left(\mathbf{k},t,m\right) = b_1\left(m,t\right)\delta\left(\mathbf{k},t\right) + 
    \frac{b_2\left(m,t\right)}{2}\int \frac{d^3\mathbf{q}}{\left(2\pi\right)^3}\delta\left(\mathbf{q},t\right)\delta\left(\mathbf{k}-\mathbf{q},t\right) + 
    \frac{b_{s_2}\left(m,t\right)}{2}\int \frac{d^3\mathbf{q}}{\left(2\pi\right)^3}\delta\left(\mathbf{q},t\right)\delta\left(\mathbf{k}-\mathbf{q},t\right)\text{S}_{2}\left(\mathbf{q},\mathbf{k} - \mathbf{q}\right),
\end{equation}
with 
\begin{equation}
\label{TidalKernel}
   \text{S}_{2}\left(\mathbf{k}_1,\mathbf{k}_2\right) = \frac{\left(\vec{k}_1\cdot\vec{k}_2\right)^2}{k_1^2k_2^2} - \frac{1}{3}, \qquad   b_{s_2}\left(m,t\right) = -\frac{2}{7}\left(b_1\left(m,t\right) - 1\right).
\end{equation}
The local biases $b_1(m,t)$ and $b_2(m,t)$ can instead be predicted via the peak-background split approach from the halo mass function $f_{\mathrm{m}}\left(m,t\right)$ . \citep{1984ApJ...284L...9K,1986ApJ...304...15B,1989MNRAS.237.1127C,1996MNRAS.282..347M}. In the following expressions we will omit the time dependence to simplify the notation.

\subsubsection{Halo model matter power spectrum}
Let us start from the power spectrum. It is well approximated at every redshift (a dependence that we will omit in the following) as the sum of the 1-halo term and of the 2-halo term which respectively captures the contribution given by the two points being in the same halo and in two separate ones. In Fourier space, the 1-halo term is simply the product of the two Fourier-transformed profiles $u\left(m,c_{\mathrm{v}},k\right)$ (we recall we use the NFW profile), convoluted with the halo mass function
\begin{equation}
 \label{P1halos}P^{1\text{h}}\left(k\right) = \int_{m^{\text{Min}}}^{m^{\text{Max}}} \mathrm{d}m\ \left(\frac{m}{\rho_{\text{com.}}}\right)^2\ f_{\mathrm{m}}\left(m\right)  \int \mathrm{d}c_{\mathrm{v}}\ p\left( c_{\mathrm{v}},m \right) u^2\left(m,c_{\mathrm{v}},k\right).
\end{equation}
In Eq.~\eqref{P1halos} $\rho_{\text{com.}}$ is the comoving background matter density and $m^{\text{Min}}, m^{\text{Max}}$ specify the integration range for the mass integral, which has to be evaluated numerically. We will comment further on how to choose these values in Sec.~\ref{subsubHMNI}.
We also accounted for the uncertainty on the halo concentration parameter via the convolution with the probability density distribution $p\left( c_{\mathrm{v}},m \right)$. It gives the probability that a virialised halo of mass $m$ has a concentration parameter $c_{\mathrm{v}}$. As anticipated, we employ the results from \cite{2001MNRAS.321..559B} who found that this distribution can be well approximated by a log-normal
\begin{equation}
\label{LogNormC}
 p\left( c_{\mathrm{v}},m \right)  = \frac{1}{c_{\mathrm{v}} \sqrt{2 \pi \sigma^2_{\ln c_{\mathrm{v}}}}}\exp\left[-\frac{\left( \ln c_{\mathrm{v}}-\ln\bar{c}_{\mathrm{v}}\left(m\right) \right)^2}{2\sigma^2_{\ln c_{\mathrm{v}}}}\right]  
\end{equation}
where $\bar{c}_{\mathrm{v}}\left(m\right)$ is the median concentration parameter for every redshift and mass and $\sigma_{\text{ln}c_{\mathrm{v}}} = 0.18$, independent from the redshift. 
The 2-halo term describes the correlation between 2 points hosted in two different halos of mass (e.g.) $m_1$ and $m_2$. Then it depends on the halo-halo correlation function. Qualitatively we write
\begin{equation}
\label{Qual2halo}
    \langle\delta_{\mathrm{h}}\left(\mathbf{k}_1,m_1\right)\delta_{\mathrm{h}}\left(\mathbf{k}_2,m_2\right)\rangle \approx b_1\left( m_1\right)b_1\left( m_2\right)\langle\delta\left(\mathbf{k}_1\right)\delta\left(\mathbf{k}_2\right)\rangle \approx b_1\left( m_1\right)b_1\left( m_2\right)P^{\text{lin.}}\left(k_1\right) \delta_{\mathrm{D}}\left(\mathbf{k}_1 + \mathbf{k}_2\right).
\end{equation}
In Eq.~\eqref{Qual2halo} we stopped the bias expansion at the linear level, i.e. $\sim (b_1\left( m_2\right)\delta)$, since it is a common assumption to consider the scales here involved (beyond the varialization halo radius) in the linear regime \citep{2001ApJ...554...56C}. For this reason, we neglect quadratic corrections and the power spectra are the linear ones. A more quantitative evaluation leads to \citep{2001ApJ...554...56C}
\begin{equation}
    \label{P2halos}P^{2\text{h}}\left(k\right) = \left[
\int_{m^{\text{Min}}}^{m^{\text{Max}}} \mathrm{d}m\ b_{1}\left(m\right) \frac{m}{\rho_{\text{com.}}}\ f_{\mathrm{m}}\left(m\right)  \left(\int \mathrm{d}c_{\mathrm{v}}\ p\left( c_{\mathrm{v}},m \right) u\left(m,c_{\mathrm{v}},k\right)\right)
\right]^2P^{\text{lin.}}\left(k\right)
\end{equation}
and the total matter power spectrum is simply the sum of the two terms derived above
\begin{equation}
\label{PspHMtot}P^{\text{HM}}\left(k\right) = P^{1\text{h}} \left(k\right) + P^{2\text{h}} \left(k\right).
\end{equation}
To simplify the expression for the general matter polyspectrum, we can introduce the following quantity
\begin{equation}
\label{ImunuHalos}
\text{I}_{\mu}^{\beta}\left(k_1, \dots, k_{\mu} \right) = \int_{m^{\text{Min}}}^{m^{\text{Max}}} \mathrm{d}m\ b_{\beta}\left(m\right) \left(\frac{m}{\rho_{\text{com.}}}\right)^{\mu}\ f_{\mathrm{m}}\left(m\right) \left(\int \mathrm{d}c_{\mathrm{v}}\ p\left( c_{\mathrm{v}},m \right) \left[\prod_{i=1}^{\mu} u\left(m,c_{\mathrm{v}},k_i\right)\right]\right)
\end{equation}
where $b_{0} \equiv 1$.
Then, the matter power spectrum~\eqref{PspHMtot} can be written in a more synthetic way as 
\begin{equation}
\label{PspHMtot2}P^{\text{HM}}\left(k\right) = \text{I}_{2}^{0}\left(k,k\right) + \left[\text{I}_{1}^{1}\left(k\right)\right]^2 P^{\text{lin.}}\left(k\right).
\end{equation}

\subsubsection{Halo model matter bispectrum}
We can write the three-dimensional matter bispectrum as the sum of 3 multi-halo terms corresponding to the following cases \citep{2001ApJ...554...56C}:
\begin{enumerate}
    \item[] $B^{1\text{h}}$, the 1-halo term: all the points are within the same halo, 
    \item[] $B^{2\text{h}}$, the 2-halo term: 2 out of 3 points are in the same halo while the third is not,
    \item[] $B^{3\text{h}}$, the 3-halo term: each point is hosted in a different halo.
\end{enumerate}
In analogy with the calculations performed at the level of power spectrum, the 1-halo term for the bispectrum is simply  the convolution of the halo mass function with the third power of the halo profile (in Fourier space)
\begin{equation}
    \label{B1halo}B^{1\text{h}}\left(k_1,k_2,k_3\right) = \int_{m^{\text{Min}}}^{m^{\text{Max}}} \mathrm{d}m\  \left(\frac{m}{\rho_{\text{com.}}}\right)^{3}\ f_{\mathrm{m}}\left(m\right) \left(\int \mathrm{d}c_{\mathrm{v}}\ p\left( c_{\mathrm{v}},m \right) \left[\prod_{i=1}^{3} u\left(m,c_{\mathrm{v}},k_i\right)\right]\right) =  \text{I}_{3}^{0}\left(k_1,k_2,k_3\right).
\end{equation}
The 2-halo term depends qualitatively on the matter density field as in Eq.~\eqref{Qual2halo} where a quantitative derivation leads to
\begin{equation}
  \label{B2halos}B^{2\text{h}} \left(k_1,k_2,k_3\right) = \text{I}_1^1\left(k_1\right)\text{I}_2^1\left(k_2,k_3\right)P^{\text{lin.}}\left(k_1\right) + 					 				 \text{I}_1^1\left(k_3\right)\text{I}_2^1\left(k_1,k_2\right)P^{\text{lin.}}\left(k_3\right) +
\text{I}_1^1\left(k_2\right)\text{I}_2^1\left(k_3,k_1\right)P^{\text{lin.}}\left(k_2\right).
\end{equation}
Finally, the 3-halo term has the following dependence
\begin{align}
\label{3haloexpansion}
\langle\delta_{\mathrm{h}}\left(\mathbf{k}_1,m_1\right)&\delta_{\mathrm{h}}\left(\mathbf{k}_2,m_2\right)\delta_{\mathrm{h}}\left(\mathbf{k}_3,m_3\right)\rangle \approx \nonumber\\
    \Big\langle
    &\left(b_1\left(m_1\right)\delta\left(\mathbf{k}_1\right) + 
    \frac{b_2\left(m_1\right)}{2} \delta\left(\mathbf{q}\right)\cdot\delta\left(\mathbf{k}_1-\mathbf{q}\right) + 
    \frac{b_{s_2}\left(m_1\right)}{2}\delta\left(\mathbf{q}\right)\cdot_{s}\delta\left(\mathbf{k}_1-\mathbf{q}\right)\right)\nonumber\\
    &\left(b_1\left(m_2\right)\delta\left(\mathbf{k}_1\right) + 
    \frac{b_2\left(m_2\right)}{2} \delta\left(\mathbf{q}\right)\cdot\delta\left(\mathbf{k}_2-\mathbf{q}\right) + 
    \frac{b_{s_2}\left(m_2\right)}{2}\delta\left(\mathbf{q}\right)\cdot_{s}\delta\left(\mathbf{k}_2-\mathbf{q}\right)\right)\nonumber\\
    &\left(b_1\left(m_3\right)\delta\left(\mathbf{k}_1\right) + 
    \frac{b_2\left(m_3\right)}{2} \delta\left(\mathbf{q}\right)\cdot\delta\left(\mathbf{k}_3-\mathbf{q}\right) + 
    \frac{b_{s_2}\left(m_3\right)}{2}\delta\left(\mathbf{q}\right)\cdot_{s}\delta\left(\mathbf{k}_3-\mathbf{q}\right)\right)\Big\rangle     
\end{align}
where we wrote in a synthetic way the convolutions~\eqref{ExpansionBiasF} via the operators $\cdot$ and $\cdot_{s}$ and $m_{1,2,3}$ are the masses of the three halos hosting the points used for computing the correlation. According to the Perturbation Theory (PT) paradigm (see \cite{2002PhR...367....1B} for a thorough review), we can expand the field $\delta\left(\mathbf{k}\right)$ at different orders
    \begin{equation}
    \label{ExpandDoublet2}
        \delta\left(\mathbf{k}\right) = \sum_p\delta^{(p)}\left(\mathbf{k}\right),
    \end{equation}
each term of the expansion being proportional to the p$^{\text{th}}$ power of the linearly-evolved initial over-density \citep{1986ApJ...311....6G,1992PhRvD..46..585M,1996ApJ...456...43J}. The leading non-vanishing term in Eq.~\eqref{3haloexpansion} includes at least one mode at second order in PT $\delta^{(2)}$. Overall, the 3-halo term at leading order is proportional to the $4^{\mathrm{th}}$ power of the initial linearly-evolved contrast density field, i.e.
\begin{equation}
\label{3haloTerm1}
    \langle\delta_{\mathrm{h}}\left(\mathbf{k}_1,m_1\right)\delta_{\mathrm{h}}\left(\mathbf{k}_2,m_2\right)\delta_{\mathrm{h}}\left(\mathbf{k}_3,m_3\right)\rangle\arrowvert_{4^{\mathrm{th}}}  \approx 
    b_1\left(m_1\right)b_1\left(m_2\right)b_1\left(m_3\right)\langle\delta^{(1)}\left(\mathbf{k}_1\right)\delta^{(1)}\left(\mathbf{k}_2\right)\delta^{(2)}\left(\mathbf{k}_3\right)\rangle + \text{cycles over}\ \{\mathbf{k}_1,\mathbf{k}_2,\mathbf{k}_3\}.
\end{equation}
The correlation (qualitatively) $\langle\delta^{(1)}\delta^{(1)}\delta^{(2)}\rangle$ can be related to the tree-level PT matter bispectrum $B^{\text{PT}}\left(k_1,k_2,k_3\right)$. 
By looking at Eq.~\eqref{3haloexpansion}, we can see that at the same order  more components are present, specifically those including $b_2\delta^2$ and $b_{s_2}\delta^2$ with $\delta$ in the linear regime, i.e. $\delta = \delta^{(1)}$. These correlations of linear modes can be reduced via the Wick theorem in products of 2-point correlation functions as 
\begin{align}
\label{3haloTerm2}
  \langle\delta_{\mathrm{h}}\left(\mathbf{k}_1,m_1\right)&\delta_{\mathrm{h}}\left(\mathbf{k}_2,m_2\right)\delta_{\mathrm{h}}\left(\mathbf{k}_3,m_3\right)\rangle\arrowvert_{4^{\mathrm{th}}} \approx \nonumber\\
  &b_1\left(m_1\right)b_1\left(m_2\right)b_{2,(s_2)}\left(m_3\right)
  \langle\delta^{(1)}\left(\mathbf{k}_1\right)\delta^{(1)}\left(\mathbf{k}_2\right)\delta^{(1)}\left(\mathbf{q}\right)\delta^{(1)}\left(\mathbf{k}_3-\mathbf{q}\right)\rangle + \text{cycles over}\ \{\mathbf{k}_1,\mathbf{k}_2,\mathbf{k}_3\} = \nonumber\\ &b_1\left(m_1\right)b_1\left(m_2\right)\frac{b_{2,(s_2)}\left(m_3\right)}{2}\Big[\langle\delta^{(1)}\left(\mathbf{k}_1\right)\delta^{(1)}\left(\mathbf{q}\right)\rangle\langle\delta^{(1)}\left(\mathbf{k}_2\right)\delta^{(1)}\left(\mathbf{k}_3-\mathbf{q}\right)\rangle + \nonumber\\ &\hspace{3.85cm}\langle\delta^{(1)}\left(\mathbf{k}_1\right)\delta^{(1)}\left(\mathbf{k}_3-\mathbf{q}\right)\rangle\langle\delta^{(1)}\left(\mathbf{k}_2\right)\delta^{(1)}\left(\mathbf{q}\right)\rangle\Big] + \text{cycles over}\ \{\mathbf{k}_1,\mathbf{k}_2,\mathbf{k}_3\} \approx \nonumber\\ &b_1\left(m_1\right)b_1\left(m_2\right)b_{2,(s_2)}\left(m_3\right)P^{\text{lin.}}\left(k_1\right)P^{\text{lin.}}\left(k_2\right) + \text{cycles over}\ \{\mathbf{k}_1,\mathbf{k}_2,\mathbf{k}_3\}
\end{align}
where we did not consider the contributions forcing $\mathbf{k}_i = \mathbf{k}_j,\ (i,j = 1,2,3)$ since they are associated to degenerate triangular configurations in the original bispectrum. A detailed evaluation of the mass integration over the halo profiles leads to 
\begin{align}
    \label{B3halos}B^{3\text{h}} &\left(k_1,k_2,k_3\right) = \prod_{i=1}^3 I_1^1\left(k_i\right)B^{\text{PT}}\left(k_1,k_2,k_3\right) + \nonumber\\ &\ \ + I_1^1\left(k_1\right)I_1^1\left(k_2\right)I_1^2\left(k_3\right)P^{\text{lin.}}\left(k_1\right)P^{\text{lin.}}\left(k_2\right) + 2\ \text{terms from cycles over}\ \{k_1,k_2,k_3\} + \nonumber\\
&\ \ + \frac{4}{7}\left[I_1^1\left(k_1\right)I_1^1\left(k_2\right)\text{S}_{2}\left(\mathbf{k}_1,\mathbf{k}_2\right)P^{\text{lin.}}\left(k_1\right)P^{\text{lin.}}\left(k_2\right)\left( 1 - I_1^1\left(k_3\right)\right) + 2\ \text{terms from cycles over}\ \{\mathbf{k}_1,\mathbf{k}_2,\mathbf{k}_3\} \right].
\end{align}
and we can recognise the tree-level PT power spectra (linear power spectra) and bispectrum induced by the correlations obtained in Eq.~\eqref{3haloTerm2} and in Eq.~\eqref{3haloTerm1}  respectively.
The total bispectrum is then the sum of the terms derived above 
\begin{equation}
\label{BspHMtot}B^{\text{HM}}\left(k_1,k_2,k_3\right) = B^{1\text{h}} \left(k_1,k_2,k_3\right)  + B^{2\text{h}} \left(k_1,k_2,k_3\right) + B^{3\text{h}} \left(k_1,k_2,k_3\right).
\end{equation}

\subsection{The halo model: matter polyspectra for the covariance matrix}
\label{CovHM}
\begin{figure}
\centering
\includegraphics[width=\hsize,clip=true]{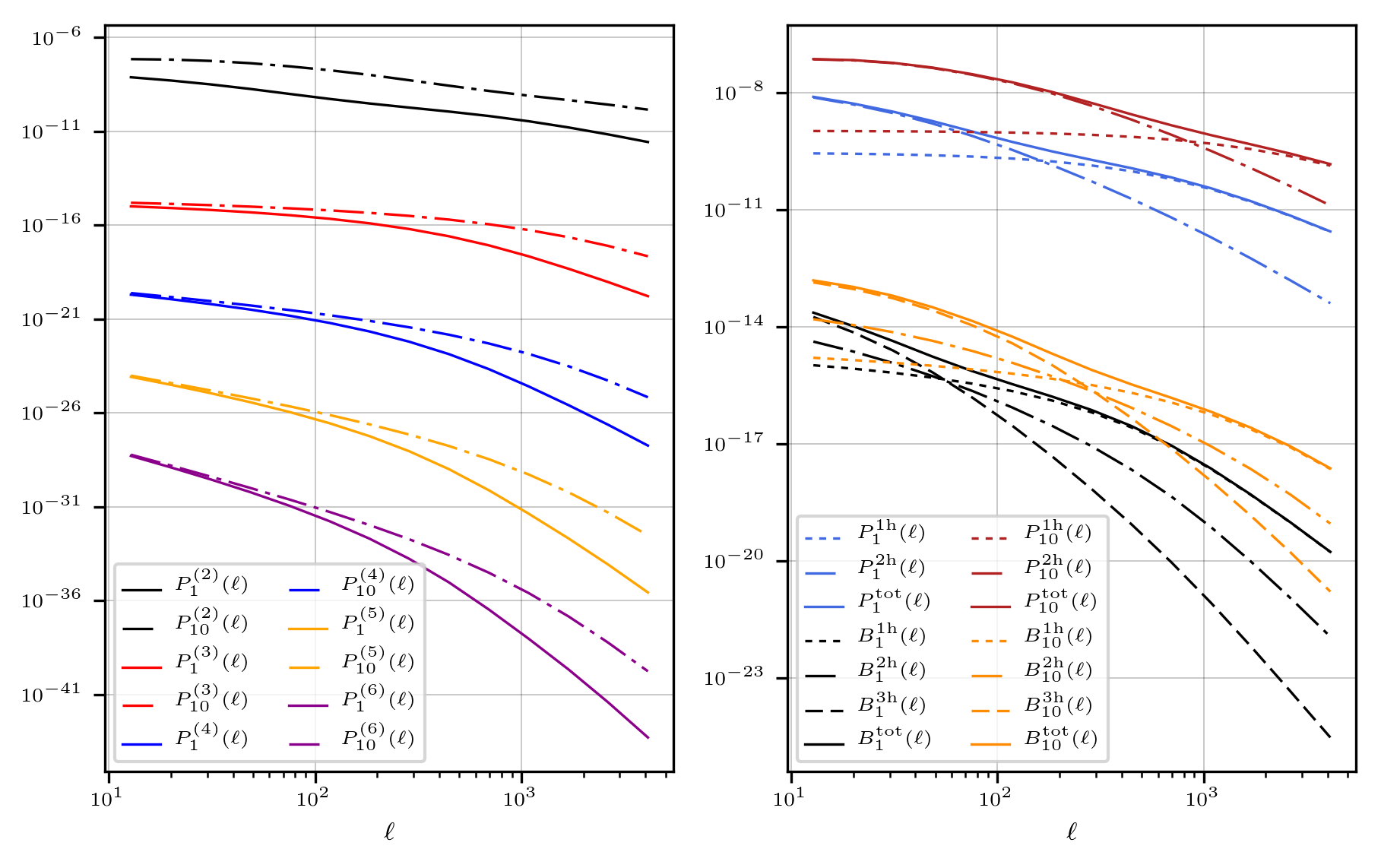}
\caption[]{\textit{Left:} auto-tomographic polyspectra of different orders (from 2 to 6) evaluated on equilateral configurations. We are considering auto-correlations for the  $1^{\text{st}}$ and the $10^{\text{th}}$ tomographic bin, in a Euclid-like photometry (Appendix~\ref{App:EuclidSpec}).  \textit{Left:} polyspectra employed in the computation of the covariance, i.e. power spectrum including the  1-+2-halo term and higher order polyspectra approximated to the 1-halo component. In the legend we are employing the notation $P^{(n)}_{i}(\ell) \equiv P_{i_1\dots i_n}\left(\ell_1,\dots,\ell_n\right)$ with $i_1 = \dots = i_n$ and $\ell = \ell_i = \dots \ell_n$, the second member of the equivalence being introduced in Eq.~\eqref{CorrFunct}.   \textit{Right}: polyspectra employed in the computation of the data vector: power spectrum including the  1-+2-halo term (we also depict the separate behaviour for each of them) and bispectrum including 1-+2-+3-halo term (we also depict the separate behaviour for each of them). In this panel we switched to the standard notation for the bispectrum, i.e. $B = P^{(3)}$.}
\label{fig:spectra}
\end{figure}

As anticipated in Sec.~\ref{sec:halomodel}, we approximate the polyspectra within the covariance matrix (beyond the 2-point correlation function) with the respective 1-halo terms. We give here their general expression for an $n$-order polyspectrum. While in real space they would require an $n$-dimensional convolution of halo profiles, the Fourier analysis makes it much simpler turning it into a one-dimensional (mass) integration of the product of $n$ Fourier-transformed halo profiles, shortly
\begin{equation}
\label{General1h}
P^{\text{1h}}\left( \mathbf{k}_1, \dots, \mathbf{k}_{n} \right) = \text{I}_{n}^{0}\left(k_1, \dots, k_{n} \right).
\end{equation}
Also, the dependence on solely the magnitude of the vectors derives from the fact that chosen single halo profile (NFW profile) is spherical in real space.
As a summary, in Fig.~\ref{fig:spectra} we show the behaviour of the polyspectra required for our analyses, both at the level of covariance (left panel) and at the level of data vector (right panel).

\subsection{The halo model: matter polyspectrum responses}
\label{subsubResp}
The halo model also provides a powerful recipe for the computation of the responses in Eq.~\eqref{CovPP22}, Eq.~\eqref{NGsscBB} and in Eq.~\eqref{NGsscBP} required for the forward modelling of the super-sample covariance. We can obtain them 
by taking the derivative of the power spectrum~\eqref{PspHMtot} and of the bispectrum~\eqref{BspHMtot} with respect to the long mode $\delta_{\mathrm{b}}$ \eqref{BackgroundMode}. We assume that the single halo profile is not affected by the long mode while it impacts the distribution of halos on larger scales via the bias and the mass function. Following \cite{2018PhRvD..97d3532C}, we consider linear responses in $\delta_{\mathrm{b}}$ and up to the linear bias (the second order ones were proved to be sub-leading in \cite{2018PhRvD..97d3532C} and we neglect for consistency the tidal component of the response being also quadratic in $\delta$). We can formally express the impact of the long mode with the help of Eq.~\eqref{ImunuHalos}
\begin{equation}
\label{ImunuHalosDeltaB}
    \frac{\partial \text{I}_{\mu}^{\beta}\left(k_1, \dots, k_{\mu} |\delta_{\mathrm{b}} \right)}{\partial\delta_{\mathrm{b}}}
 = \int_{m^{\text{Min}}}^{m^{\text{Max}}} \mathrm{d}m\  \left(\frac{m}{\rho_{\mathrm{com.}}}\right)^{\mu}  \frac{\partial }{\partial\delta_{\mathrm{b}}}\left[b_{\beta}\left(m\right)\ f_{\mathrm{m}}\left(m\right) \right]\left(\int \mathrm{d}c_{\mathrm{v}}\ p\left( c_{\mathrm{v}},m \right) \left[\prod_{i=1}^{\mu} u\left(m,c_{\mathrm{v}},k_i\right)\right]\right) = \text{I}_{\mu+1}^{\beta}\left(k_1, \dots, k_{\mu} \right)
\end{equation}
where the derivative appearing within the expression above can be computed via the peak-background split approach as \citep{1974ApJ...187..425P,1989MNRAS.237.1127C,1996MNRAS.282..347M,1999MNRAS.308..119S,2018PhR...733....1D}
\begin{equation}
    b_{\beta}\left(m\right) = \frac{1}{f_{\mathrm{m}}\left(m\right)} \frac{\partial f_{\mathrm{m}}\left(m\right)}{\partial\delta_{\mathrm{b}}}.
\end{equation}
Therefore, the halo model power spectrum linear response is related to the one of the linear matter power spectrum as 
\begin{equation}
\label{PowerHMResp}\frac{\mathrm{d}P^{\text{HM}}\left(k|\delta_{\mathrm{b}}\right)}{\mathrm{d} \delta_{\mathrm{b}}}\Big\arrowvert_{\delta_{\mathrm{b}}=0} \approx \left[\text{I}^1_1\left(k\right)\right]^2\frac{\mathrm{d} P^{\text{lin.}}\left(k|\delta_{\mathrm{b}}\right)}{\mathrm{d} \delta_{\mathrm{b}}}\Big\arrowvert_{\delta_{\mathrm{b}}=0} + \text{I}^1_2\left(k\right).
\end{equation}
In Eq.~\eqref{PowerHMResp} we neglected the term proportional to $\text{I}_2^2$ begin smaller than those already included \citep{2014JCAP...05..048C,2015JCAP...08..042W,2018PhRvD..97d3532C}.
Under the same approximations, the halo model bispectrum linear response can be expressed as 
\begin{align}
\label{BispHMResp}\frac{\mathrm{d}B^{\text{HM}}\left(k_1,k_2,k_3\right|\delta_{\mathrm{b}})}{\mathrm{d} \delta_{\mathrm{b}}}\Big\arrowvert_{\delta_{\mathrm{b}}=0} &\approx \text{I}^1_1\left(k_1\right)\text{I}^1_1\left(k_2\right)\text{I}^1_1\left(k_3\right)\frac{\mathrm{d} B^{\text{PT}}\left(k_1,k_2,k_3\right|\delta_{\mathrm{b}})}{\mathrm{d} \delta_{\mathrm{b}}}\Big\arrowvert_{\delta_{\mathrm{b}}=0} + \nonumber\\
&+ \left[ \text{I}^1_1\left(k_1\right)\text{I}^2_2\left(k_2,k_3\right)P^{\text{lin.}}\left(k\right) + \text{I}^1_1\left(k_1\right)\text{I}^1_2\left(k_2,k_3\right)\frac{\mathrm{d} P^{\text{lin.}}\left(k_1|\delta_{\mathrm{b}}\right)}{\mathrm{d} \delta_{\mathrm{b}}}\Big\arrowvert_{\delta_{\mathrm{b}}=0}\right] +\ 2\ \text{cycles}\ + \nonumber\\
&+ \text{I}^1_3\left(k_1,k_2,k_3\right).
\end{align}
Eq.~\eqref{PowerHMResp} and Eq.~\eqref{BispHMResp} require the evaluation at tree-level in perturbation theory of the linear response for the power spectrum and for the bispectrum. The derivation is quite technical and we refer to \cite{2018PhRvD..97d3532C} for a detailed step-by-step explanation
\begin{equation}
    \frac{\mathrm{d} P^{\text{lin.}}\left(k|\delta_{\mathrm{b}}\right)}{\mathrm{d} \delta_{\mathrm{b}}}\Big\arrowvert_{\delta_{\mathrm{b}}=0} \approx \frac{47}{21} P^{\text{lin.}}\left(k\right) - \frac{1}{3} \frac{\mathrm{d}P^{\text{lin.}}\left(k\right)}{\mathrm{d}\ln k},\label{RespP}
\end{equation}
\begin{equation}
\frac{\mathrm{d} B^{\text{PT}}\left(k_1,k_2,k_3|\delta_{\mathrm{b}}\right)}{\mathrm{d} \delta_{\mathrm{b}}}\Big\arrowvert_{\delta_{\mathrm{b}}=0} \approx \frac{433}{126}B^{\text{PT}}\left(k_1,k_2,k_3\right) + \frac{5}{126}B_{G_2}\left(k_1,k_2,k_3\right) -\frac{1}{3}\sum_{i=1}^3 \frac{\mathrm{d} B^{\text{PT}}\left(k_1,k_2,k_3\right)}{\mathrm{d}\ln k_i}.\label{RespB}
\end{equation}
By replacing the responses \eqref{RespP} and \eqref{RespB} into Eqs.~\eqref{PowerHMResp} and \eqref{BispHMResp}, we obtain the desired expressions.

\subsection{Halo model numerical implementation}
\label{subsubHMNI}
We would like to conclude this section by discussing more on a particular point: the mass numerical integration within Eq.~\eqref{ImunuHalos}. The main consequence of having a numerical  integration over a finite domain is the exclusion of halos whose masses fall outside the range $\left(m^{\text{Min}},m^{\text{Max}}\right)$. 
This cut-off is intrinsically present in all known halo model implementations given that they are calibrated against simulations over a finite mass range. However, we do not expect this feature to have any impacts: very light halos, extending over extremely small scales, should not contribute to our cosmological observables while heavy halos are exponentially suppressed due to the shape of the mass function.
The sensitivity to the mass cut-off is in particular problematic for the lower bound. We can easily see this point by looking at the integrand $\mathcal{I}_{\beta}(m)$ in the following consistency relations \citep{2002PhR...372....1C,10.1046/j.1365-8711.2003.06321.x}
\begin{equation}
\label{BoundinMass}
\int_{m^{\text{Min}}}^{m^{\text{Max}}} \frac{m}{\rho_{\text{com.}}} \mathrm{d}m\ b_{\beta}\left(m\right) \ f_{\mathrm{m}}\left(m\right) \equiv\ \int_{m^{\text{Min}}}^{m^{\text{Max}}}\mathcal{I}_{\beta}(m)\ \mathrm{d}m \ =\  \begin{cases}
						   1 & \mr{if} \ \beta \le 1,\\
                           0 & \mr{if} \ \beta \ge 2,
						   \end{cases} 
\end{equation}
where the case for $\beta=0$ is actually the consistency relation for the mass function. In physical terms, the relations Eq.~\eqref{BoundinMass} requires that the mass of the Universe is entirely enclosed in halos ($\beta=0$) of mass $m\in\left(m^{\text{Min}},m^{\text{Max}}\right)$ and that the overall distribution of halos is not biased compared to the total matter distribution ($\beta > 0$), at each order in the bias expansion. 
\begin{figure}
\centering
\includegraphics[width=\hsize,clip=true]{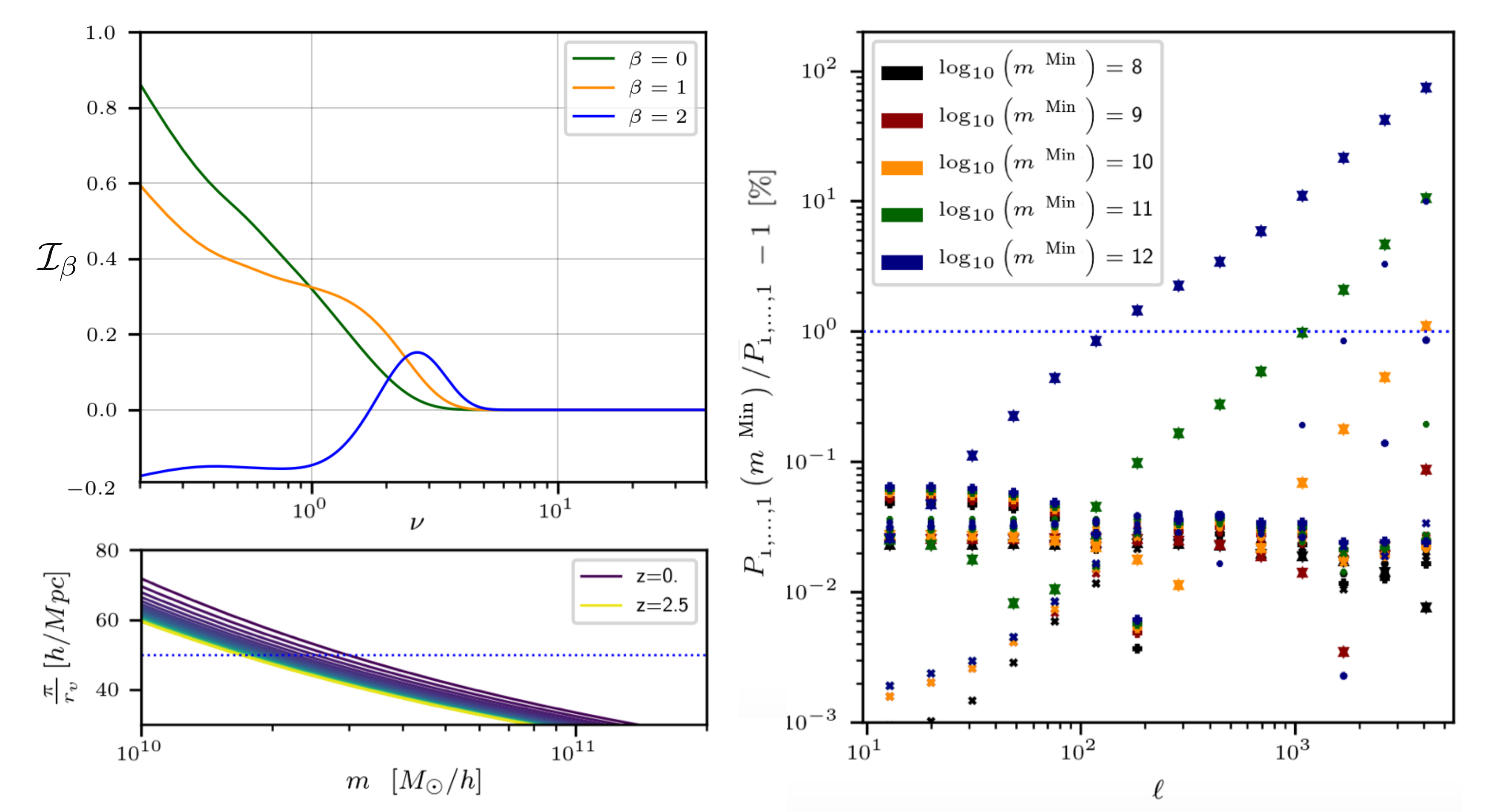}
\caption[]{\textit{Top-left:} integrand $\mathcal{I}_{\beta}$ in Eq.~\eqref{ImunuHalos} plotted against $\nu$ (see main text for definition) for $\beta= 0, 1, 2$.  \textit{Bottom-left:} inverse halo comoving radius as function of the virialised mass at different redshift. \textit{Right:} impact on the polyspectra, from the order 2 to 6, from excluding light halos from the mass integrations required in their computation. The minimum mass considered in the different cases is given in the legend (and included as further dependence on the spectra) and we are not differentiating polyspectra of different orders since interested in the overall behaviour. We are considering auto-tomographic polyspectra for the first tomographic bin evaluated on equilateral configuration $P_{1,\dots,1}(\ell,\dots,\ell)$ (on the y-axis label we are omitting the dependence on the multipoles) and we are including a dependence on the lower integration bound $m^{\mathrm{Min}}$ employed. We investigate the sensitivness of the polyspectra to the low-mass cut-off by computing the percentage fractional difference with a conservative case $\bar{P}_{1,\dots,1}(\ell,\dots,\ell)$, derived by integrating over halos of mass $m \geq 10^2 \Msun/h$.}
\label{fig:BiasMass}
\end{figure}
In the top-left panel of Fig.~\ref{fig:BiasMass}, we plot the integrand $\mathcal{I}_{\beta}(m)$ defined in Eq.~\eqref{BoundinMass} for $\beta=0,1,2$ as function of the variable 
\begin{equation}
    \nu\left(m,z\right) \equiv \left(\delta^{\mathrm{c}}_{\text{sp.}}\left(z\right)/\sigma\left(m\right)\right)^2.
\end{equation}
$\delta^{\mathrm{c}}_{\text{sp.}}$ and  $\sigma^2\left(m\right)$ are respectively the barrier for the spherical collapse \citep{1974ApJ...187..425P} and the variance of the linear matter field smoothed over spheres enclosing the mass $m$.
We see that the integrals we are trying to evaluate are slowly convergent for $\nu \rightarrow 0$ ($m \rightarrow 0$ implies $\sigma\left(m\right) \rightarrow \infty$). Therefore we would expect every halo model implementation to be very sensitive to extremely light halos, which contradicts the physical intuition described above and consequently undermines halo models themselves, not being testable at these small masses. The problem comes from the extrapolation of the model beyond its regime of validity assuming that arbitrarily small mass halos are present in our simulation accounting for the whole matter content and biases. Clearly Eq.~\eqref{BoundinMass} would not be satisfied since we are excluding these halos from the integration. However, once a minimum mass is set, lighter halos are excluded while the total matter budget may be filled by non-virialized dust. This contribution cannot be caught by the mass function $f_{\mathrm{m}}\left(m\right)$, by definition. In order to fulfil the normalisation constraints~\eqref{BoundinMass} we simply assume that the non-virialised matter content provides an effective contribution via the following regularisation of the mass function and of the biases \citep{2016PhRvD..93f3512S}
\begin{align}
\label{effectiveMasBias}
f_{\mathrm{m}}\left(m\right)\ \rightarrow& \ f_{\mathrm{m}}\left(m\right) + \alpha_0 \ \delta_{\text{D}}\left( m - m^{\text{Min}}\right),\\
b_{\beta}\left(m\right) \ \rightarrow& \begin{cases}
						   b_{\beta}\left(m\right) & \mr{if} \ m > m^{\text{Min}},\\
                           \alpha_{\beta} & \mr{if} \ m = m^{\text{Min}}.
						   \end{cases}
\end{align}
The parameters $\alpha_{\beta}$ $(\alpha_0)$ then are fixed in order to satisfy the consistency relations~\eqref{BoundinMass}. We do not consider any corrections related to the upper bound $m^{\text{Max}}$ because heavy halos are strongly suppressed by the mass function itself, as we can see from the  top-left panel in Fig.~\ref{fig:BiasMass}. Our numerical integrals converge for $m^{\text{Max}}\approx 10^{16}\Msun/\mr{h}$. 
In the computation of the weak lensing observables, we pushed our line-of-sight integration up to $k_{\text{Max}}=50 h\ \mathrm{Mpc}^{-1}$ \citep{2011MNRAS.416.1717K}. Therefore, we do not expect halos whose comoving radius is smaller than this scale to be significant for our analyses. In the bottom-left panel in Fig.~\ref{fig:BiasMass} we show the (inverse) comoving halo virialization radius as function of their virialised mass. The radius corresponding to the above scale of $k_{\text{Max}}=50 h\ \mathrm{Mpc}^{-1}$ (indicated by the horizontal blue dotted line) encloses a mass $ m^{k_{\text{Max}}} \approx 10^{10} \Msun/h$ (with a small dependence on the redshift). Therefore, if our set-up (once applied the regularisation \eqref{effectiveMasBias}) is consistent, our polyspectra should not be sensitive to halos lighter than $m^{k_{\text{Max}}}$. This is indeed what we prove with the right panel in Fig.~\ref{fig:BiasMass}.
Here, we look at the impact on the polyspectra, from the order 2 to 6, from excluding light halos from the mass integrations required in their computation. The minimum mass considered in the different cases is given in the legend and we are not differentiating polyspectra of different orders since interested in the overall behaviour. Also, we are presenting this analysis with single points in order to have a cleaner figure. We are considering auto-tomographic polyspectra for the first tomographic bin (being the ones more sensitive to light halos) evaluated on equilateral configuration (for simplicity), i.e. $P_{1,\dots,1}(\ell,\dots,\ell)$ (we are omitting the dependence on the multipoles on the y-axis label) according to the notation given in Eq.~\eqref{CorrFunct}. We investigate the sensitivness of the polyspectra to the low-mass cut-off by comparing them with a conservative case $\bar{P}_{1,\dots,1}(\ell,\dots,\ell)$ which is derived by integrating over halos of mass $m \geq 10^2 \Msun/h$. Our implementation is solid: we observe a first deviation by more than 1\% when integrating over halos of mass $m \gtrsim 10^{10}\Msun/h = m^{k_{\text{Max}}}$. We chose for the mass integrals $ m^{\text{Min}} = 10^9 \Msun/h$ to avoid numerical inaccuracies with a negligible extra computational price. This is extremely interesting also because this resolution in mass can be easily achieved by present smoothed-particle hydrodynamics simulations \citep{2014MNRAS.445..581H,2014MNRAS.445..175G,2015MNRAS.450.1349K} allowing precise tests of the statistical properties of the halo model at an accuracy which is sufficient for the joint power spectrum-bispectrum  analysis. In our specific case, we employed a halo model implementation based on the work of~\cite{2001MNRAS.321..559B}. In their work, the distribution of halos was tested at the level of mass function with a simulation covering a mass range $10^{11}\Msun/h - 10^{14}\Msun/h$, from redshift $z=40$ to the present.

\section{Analysis of the Information content in the weak lensing observables}
\label{sec:analysis}
\begin{figure}
\centering
\includegraphics[width=\hsize,clip=true]{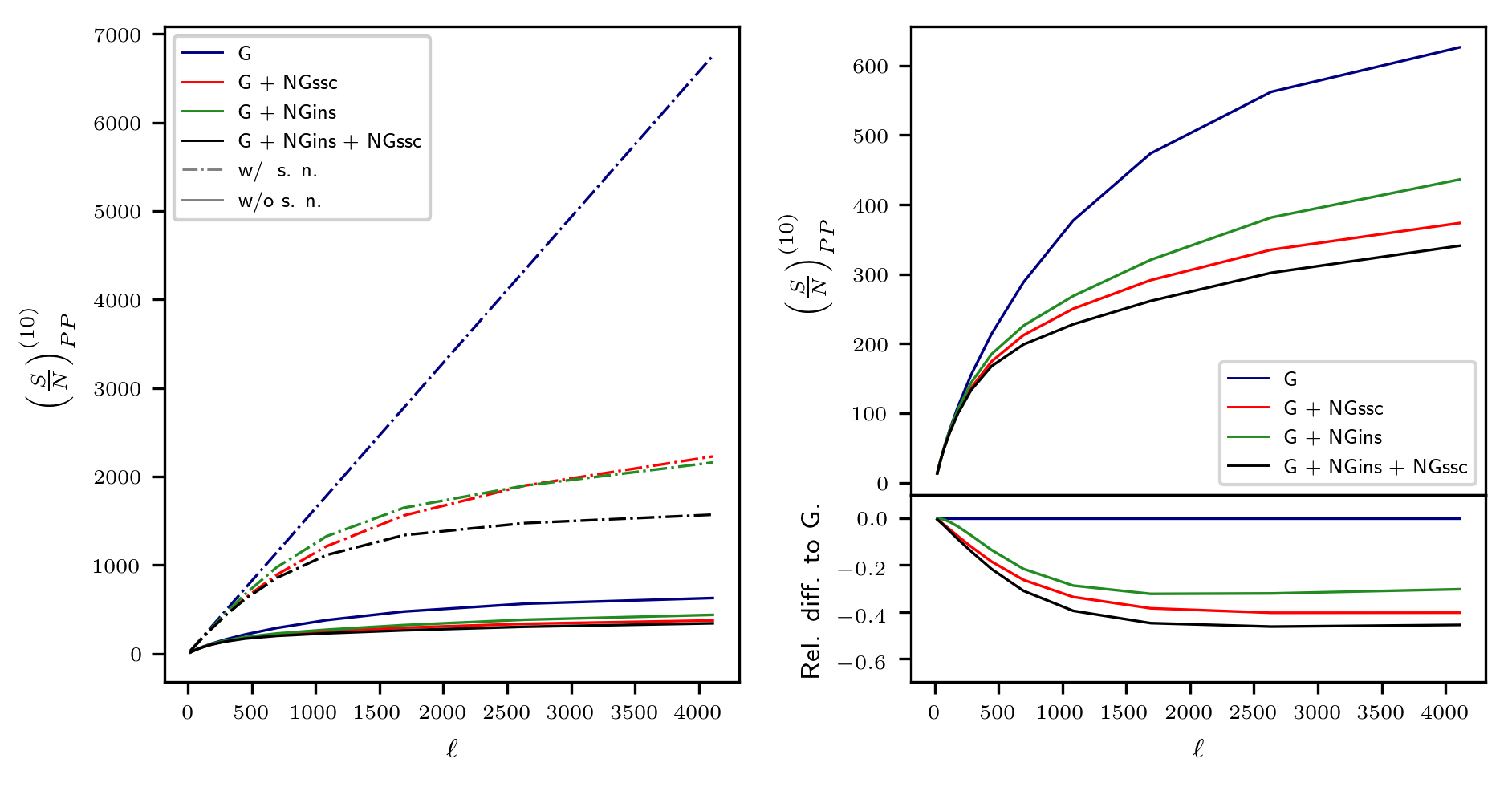}
\caption[]{\textit{Left:} S/N analysis for the power spectrum of the weak lensing convergence field when considering different contributions to the covariance matrix and the full power from a 10 bin Euclid-like tomography (as specified by the superscripts on the y-axis labels). To label them, we follow the notation given in Eq.~\eqref{CovPPsynt}. For every case, we plot 1) the  signal assuming our observations not being contaminated by shot noise (dashed lines) and 2) the signal when the shot noise is considered (solid lines). \textit{Top-right:} S/N analysis when the shot noise is considered, for different approximations to the covariance matrix. \textit{Bottom-right:} fractional differences between the S/N (shot noise included) for a purely Gaussian covariance and the S/N as computed from different approximations to the covariance.
}
\label{fig:SNPP}
\end{figure}

The major result of the work presented in this paper is the computation of remarkably large covariance matrices involved in the exploitation of the joint tomographic weak lensing convergence power spectrum-bispectrum probe and the design of an algorithm which will be a key element for future galaxy surveys. In order to deeply understand the actual benefit in terms of cosmological parameter estimation,  Fisher forecast \citep{tegmark1997measuring}, DALI forecast~\citep{2014MNRAS.441.1831S,2015MNRAS.453..893S} or Markov Chain Monte Carlo (MCMC) analyses are clearly advised: we leave these ambitious steps for future works. We underline that the covariance matrix computation is a key step in these directions and this work already finalised this calculation.
Still, we would like to have an idea of the information content achievable via the joint probe and have insights on the level of correlations between the observables.
Following~\cite{2013arXiv1306.4684K}, we define the information content of an observable as the expected inverse variance of its amplitude $A$ from a set of measured values $\vec{x}$, assuming a fixed shape. In other terms, the second power of the signal-to-noise ratio (S/N) is the inverse of the unique element of the Fisher information matrix in such a single parameter experiment. Specifically, we can write
\begin{equation}
\label{SNprim}
\left(\frac{S}{N}\right)^2 \equiv \frac{\partial\left(\vec{x}-A\vec{D}\right)^{t}}{\partial A}\cdot \text{Cov}^{-1}\cdot\frac{\partial\left(\vec{x}-A\vec{D}\right)}{\partial A} = \vec{D}^{t}\cdot \text{Cov}^{-1}\cdot\vec{D},
\end{equation}
where $A\vec{D} = \langle\vec{x}\rangle$ and $\text{Cov}_{ij}=\langle(x_i-\langle x_i\rangle)(x_j-\langle x_j\rangle)\rangle$.\\
For a joint analysis of the binned tomographic power spectrum and bispectrum, we modify Eq.~\eqref{SNprim} to account for correlations up to a maximum binned measured angular multipole $\ell_{\text{max}}$. The S/N ratio as cumulative function of the maximum multipole included can be written as
\begin{equation}
\label{eq:SN}
\left(\frac{S}{N}\right)^2_{P+B} = \sum^{i,j}_{\ell(i),\ell(j) < \ell_{\text{max}}} \text{D}_i \left[ C^D \right]^{-1}_{ij} \text{D}_j,
\end{equation}
where we need 1) a vector of observables and 2) their inverse covariance matrices
\begin{align}
\label{eq:SN2}
\vec{D} &= \{\vec{P},\ \vec{B}\},\\
C^{D} &=  \begin{pmatrix}
	    	C^{PP}&C^{PB}\\
      		C^{PB}&C^{BB}
            \end{pmatrix}.
\end{align}
The matrices $C^{PP}$, $C^{BB}$ and $C^{PB}$, are respectively the covariance of the power spectrum, bispectrum and the cross-covariance between the two as defined in Eqs.~\eqref{CovPPsynt}-\eqref{CovPBsynt}. The spectra are ordered within the vector $\vec{D}$ as suggested in \cite{2013arXiv1306.4684K}. Specifically, the power spectra evaluated on the $i^{\text{th}}$ $\ell$-bin are placed for increasing value of $i$. For each multipole, the tomographic indexes $\left(i,j\right)$ are ordered such that $i\le j$, $j$ being the faster varying index through the vector. The bispectra, evaluated over the binned configuration $\left(\ell_i, \ell_j, \ell_k \right)$, are placed in the order which satisfies $\ell_i\le\ell_j\le\ell_k$ where the index $j$ is the fastest and the index $k$ is the slowest varying one while moving along the vector. 
For each triangular configuration, the bispectra associated to the tomographic bins $\left(i,j,k\right)$ are ordered such that $k$ is the fastest index and $i$ the slowest index. While not imposing any constraint on these index for scalene triangles, we exploit symmetries at the level of triangular configurations to neglect some tomographic combinations which might eventually lead to double counting the information. We refer for more details to~\cite{2013arXiv1306.4684K}. In both cases, tomographic spectra for the same binned $\ell$-configuration are contiguous.  
When interested in accessing the cosmological information in the single observable, it is just matter of assuming $\vec{D} = \vec{P}$ (resp. $\vec{B}$) and $C^{D} = C^{PP}$ (resp. $C^{BB}$).
Given that one of the main problems for future weak lensing surveys is the modeling of the non-linear scales, we exploit the S/N 1) to understand how well the overall parameter space is constrained by our observations up to a given angular scale $\ell_{\text{max}}$, 2) to test how much the uncertainties of our theoretical model can degrade the information on the cosmological parameters and 3) to understand the impact of different approximations to the covariance matrices. On top of that, motivated by the effort in simplifying the complexity of the covariance matrices, we use this quantifier to 4) verify the robustness of simpler statistics when compared with the most complete analysis.
We would like to remind that the formalism presented so far is general and provide the theoretical background for forecasting the information content of the observables from any galaxy survey. The results showed in the following were specifically produced implementing the specificities for a Euclid-like survey described in Appendix~\ref{App:EuclidSpec}.
\begin{figure}
\centering
\includegraphics[width=\hsize,clip=true]{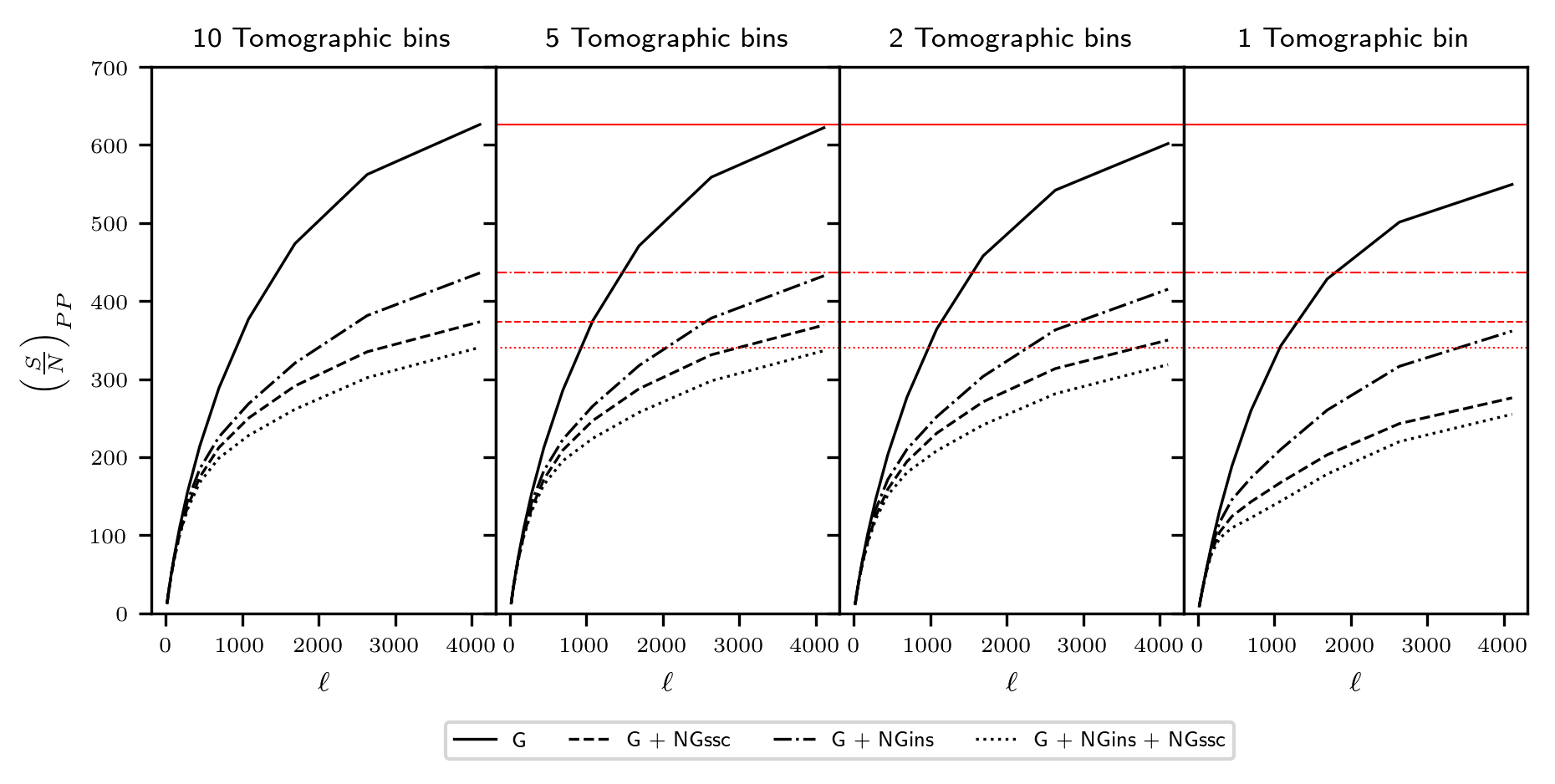}
\caption[]{S/N analysis for the power spectrum of weak lensing convergence when considering different approximations to the covariance matrix. For them, we follow the notation introduced in Eq.~\eqref{CovPPsynt}. Also, we allow for different binning in redshift of the expected sources. The number of bins indicated on the title to each panel are constructed as equipopulated, starting from the survey specificities listed in Appendix~\ref{App:EuclidSpec}. We give the edges for the different binning choices in the main text. As for the angular multipole, we keep our usual binning in $\log \ell$ (14 regularly spaced bins from $\ell=10$ to $\ell=5000$) and all the survey related parameters as chosen in Appendix~\ref{App:EuclidSpec}.
}
\label{fig:SNPPdep}
\end{figure}

\begin{figure}
\centering
\includegraphics[width=\hsize,clip=true]{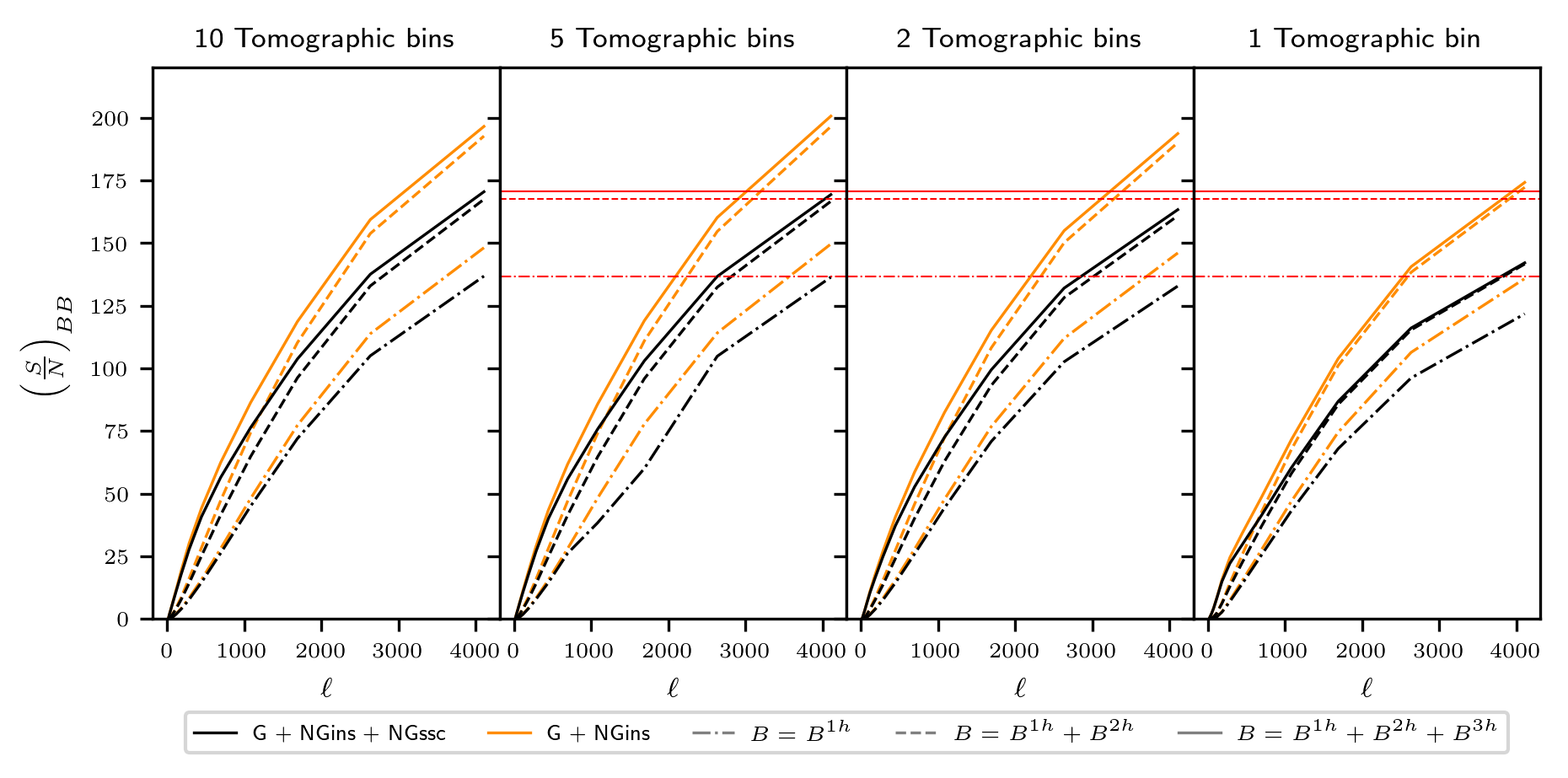}
\caption[]{S/N analysis for the bispectrum alone of weak lensing convergence when considering different approximations to the covariance matrix. We are here inspecting different effects. As in Fig.~\ref{fig:SNPPdep}, we are looking 1) at the impact of different approximations to the covariance matrix and 2) at the behaviour of the information content as function of the binning of the sources in redshift. For the first point, we label the different components of the covariance matrix according to the notation introduced in Eq.~\eqref{CovBBsynt}. On top of that, we study the impact of approximating the data vector $\vec{B}$ by neglecting different multi-halo configurations. In this analysis, we employ the same binning in $\ell$ and in redshift as for the respective panels in Fig.~\ref{fig:SNPPdep}. The shot noise is included in all the forecasts.}
\label{fig:SNBBdep}
\end{figure}
\subsection{Power spectrum signal-to-noise ratio}
In Fig.~\ref{fig:SNPP} we start our analysis from the simplest case: the power spectrum. For this first analysis we are keeping the redshift binning for the sources as in a Euclid-like tomography, i.e. 10 equi-populated bins (see Appendix~\ref{App:EuclidSpec}). We can see that, starting from a Gaussian covariance not contaminated by shot noise (dashed violet line), the main impact on the maximum information content is produced by accounting for the shot noise which degrades the S/N on all the scales. This is expected being a scale-independent contributions to the errors. We then compare the S/N as reconstructed from different approximations to the covariance, with and without the shot noise. We follow the notation introduced in Eq.~\eqref{CovPPsynt} labelling as G, NGins and NGssc respectively the Gaussian, the intra-survey and the super-survey component of the covariance.
In both cases, and with respect to the Gaussian approximation, the biggest loss of information is induced by the super-sample covariance NGssc. Focusing on the analysis including the shot noise (the more realistic case), the NGssc induces a 40\% reduction on the maximum S/N. The in-survey component NGins (alone) leads to a degradation of about 30\%. Finally, the combined effect of the two leads to a loss of about 45 \%.

In Fig.~\ref{fig:SNPPdep} we investigate at the level of power spectrum a first way to compress our data by looking at the behaviour of the maximum information content depending on the redshift binning of the sources. From the original 10 Euclid-like redshift bins, we test a first possible compression by re-binning the sources into respectively 5 (redshift bin edges: 0.001, 0.560, 0.789, 1.019, 1.324, 2.500), 2 (redshift bin edges: 0.001, 0.900, 2.500) and 1 (redshift bin edges: 0.001, 2.500) broader equipopulated intervals. In the case of the power spectrum analysis our data vector has dimension 770, 210, 42 and 14 when considering 10, 5, 2 or 1 tomographic bins respectively. The edges of the bins involved are given in the caption to Fig.~\ref{fig:SNPPdep}. For all the possible approximations to the covariance, we see that the maximum information content decreases (as expected) while moving to smaller numbers of bins. Also, the way the cumulative S/N is affected by the binning is independent on which covariance approximation is considered. For the most realistic case where no approximations on the covariance are assumed, the maximum S/N in the three situations proposed in Fig.~\ref{fig:SNPP} (from left to the right), is reduced by roughly 1\%, 5\% and 25\% when compared to the 10 bin case, whose maximum value is represented by red lines within the 3 panels on the right. 
Therefore, this analysis shows that a forecast based on a 5 bin tomography does not spoil our knowledge on the cosmological parameters for more than 1 \% while simplifying a lot our computation.  Indeed, the vector of power spectra in the 5 and 10 bin cases is made of 210 and 770 elements respectively, so that the 5 bins case gives a $\sim 3.7$ reduction of the data vector size and a $\sim 13.4$ reduction of the covariance matrix size. 

\subsection{Bispectrum signal-to-noise ratio}
In Fig.~\ref{fig:SNBBdep} we show for the bispectrum alone a similar analysis as the one proposed in Fig.~\ref{fig:SNPPdep} for the power spectrum. On top of that, we are also considering different approximations to the bisepctra in the data vector $\vec{B}$, namely excluding different multi-halo contributions to its description. By comparing the black and yellow solid lines (no approximations made on the data vector), we can inspect instead the impact of different approximations to the covariance matrix. Also, we investigate the possibility to compress the data vector via re-binning of the sources.

Starting from the possibility of data compression, we can see that a forecast based on 5 equipopulated redshift bins still allows us to recover the full information content for a Euclid-like survey. In the case of the bispectrum, the data vector $\vec{B}$ has dimension 72280, 9790, 776 and 130 when considering 10, 5, 2 or 1 tomographic bins respectively. The 5 bin tomography then allows us for a reduction of the size of the data vector and of the covariance matrix by a factor $\sim 7.4$ and $\sim 54.8$  respectively. The edges of the bins involved are given in the caption to Fig.~\ref{fig:SNPPdep}

Secondly, we analyse the impact on the maximum S/N of different approximations both at the level of covariance (while not having any approximations on the data vector $\vec{B}$ - solid lines) and at the level of data vector $\vec{B}$ (while keeping the same covariance - lines of the same colour).
Independently of the tomographic analysis, neglecting the 3-halo term leads to a very small effect on the information content at all scales ($\sim 1$\%). The analysis is instead more sensitive to the 2-halo term. 
The super-sample covariance clearly leads to a major degradation of the information content in the bispectrum. By comparing the analyses when all the multi-halo configurations are included in the bispectra (solid lines),  we can see that the super-sample covariance reduces the information content by $\sim$ 20\% for the case of 1 bin tomography, $\sim$ 15\% for the 2 and 5 bins cases and of $\sim$ 13\% for the 10 bins tomography.\footnote{The results presented in this section represent a novelty with respect to the first version of the article. In the first version, due to a numerical error in the computation of the bispectrum super-sample covariance, we erroneously detected no effect on the signal-to-noise ratio from these extra correlations. After correcting for it, we find that the super-sample covariance can not be removed from the analysis.}.

\begin{figure}
\centering
\includegraphics[width=\hsize,clip=true]{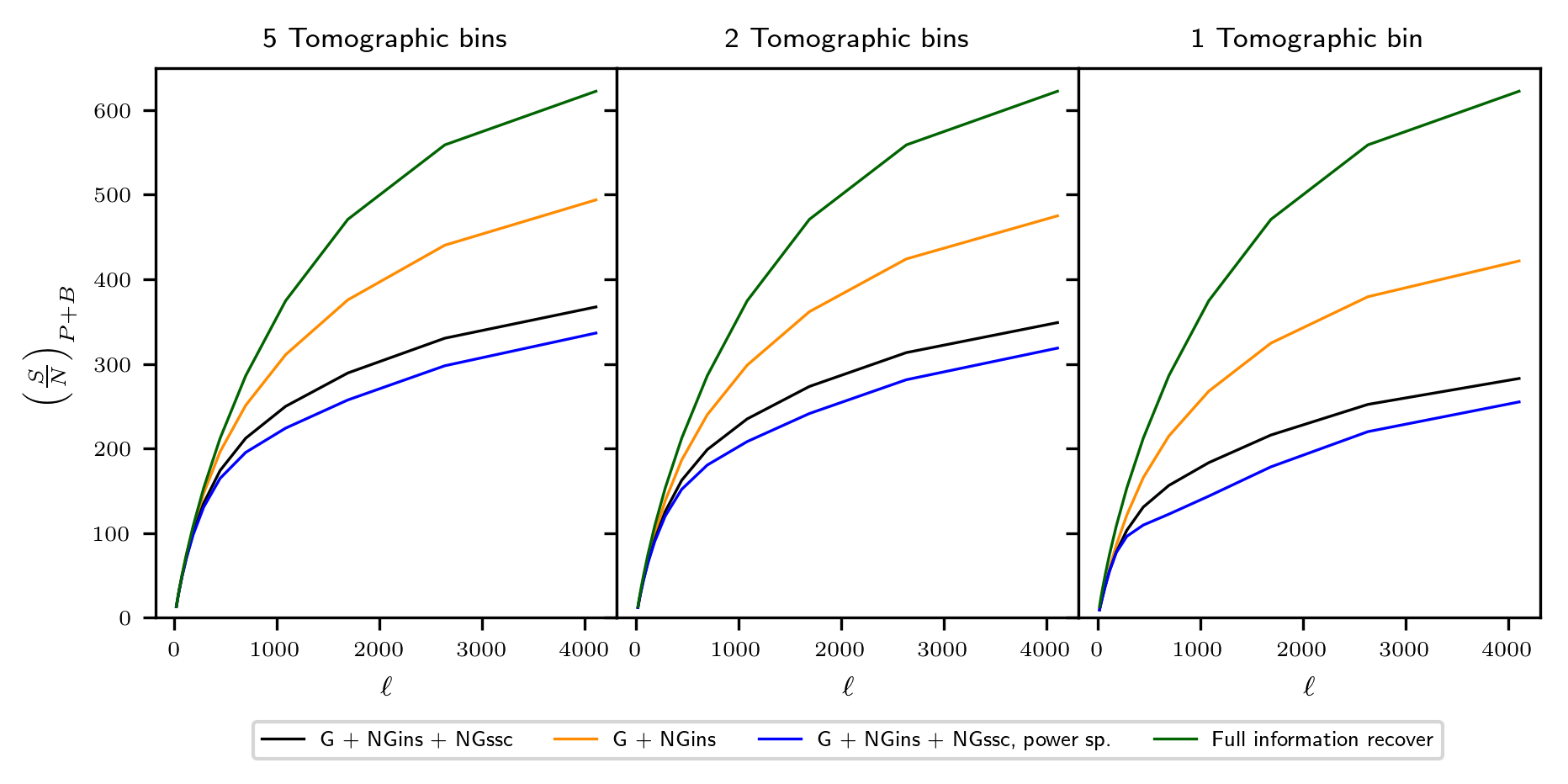}
\caption[]{S/N analysis for the weak lensing convergence power spectrum-bispectrum joint probe. We compare the analyses when including (black solid line) and excluding (yellow solid line) the NGssc component. The G and NGins components are kept in both cases. We are including all the required multi-halo configurations in the bispectrum data vector $\vec{B}$. As in Fig.~\ref{fig:SNPPdep}, we are also focusing on the behaviour of the information content as function of the binning of the sources in redshift. We employ the same bins in $\ell$ and the same survey properties as in the three respective panels in Fig.~\ref{fig:SNPPdep} and in Fig.~\ref{fig:SNBBdep}. In all the three panels we also depict the cumulative S/N for the power spectrum probe, when including all the components to the associated covariance (blue solid line). The green solid line represent the hypothetical case of a pure Gaussian weak lensing convergence field contaminated with shot noise.}
\label{fig:SNFULLdep}
\end{figure}
\begin{figure}
\centering
\includegraphics[width=\hsize,clip=true]{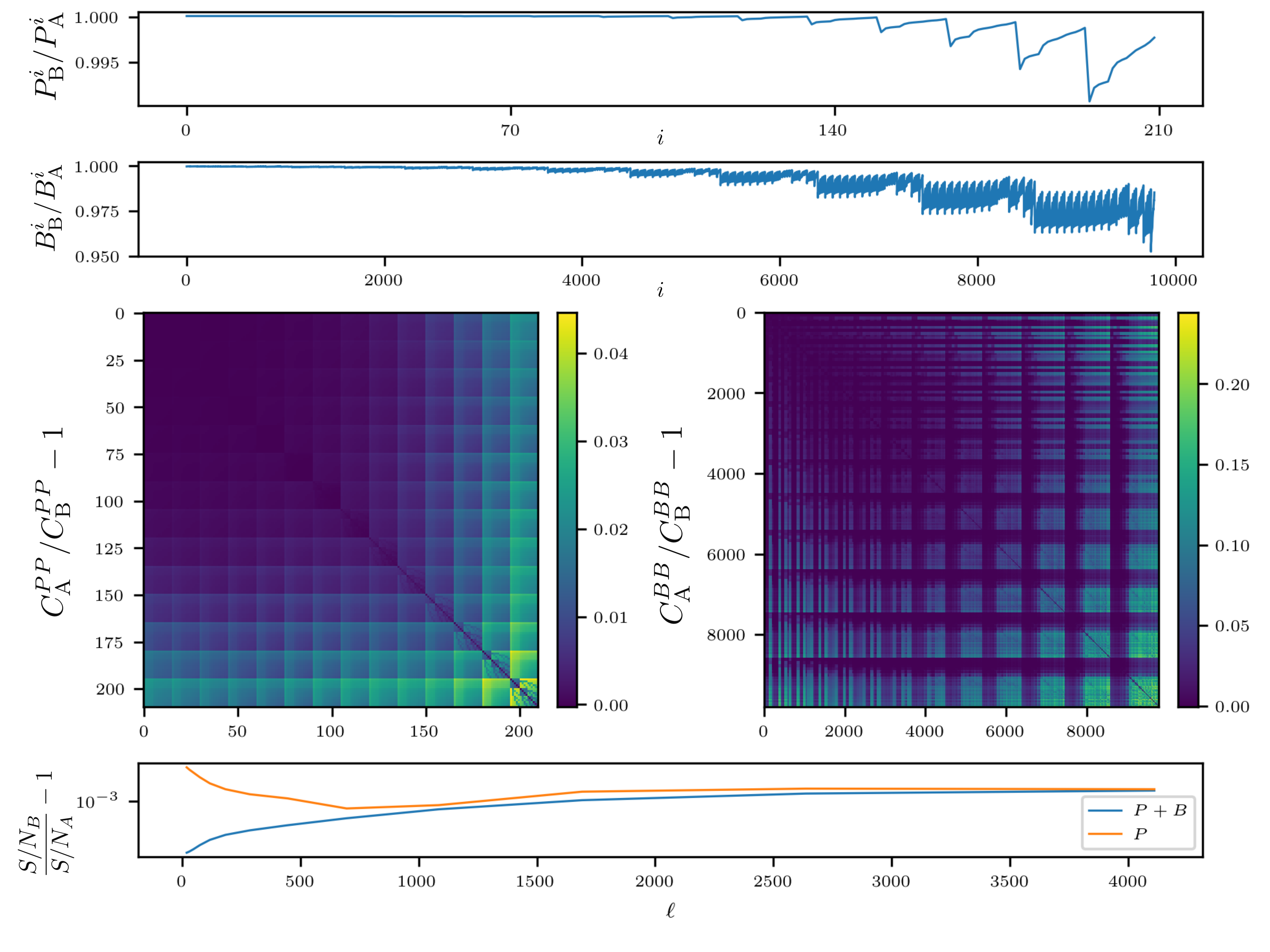}
\caption[]{Fractional differences between observables (data vector and covariance) as derived from the model A and B (defined in the main text). From the top to the bottom: 1) vector of power spectra $\vec{P}$, 2) vector of bispectra $\vec{B}$, 3 (left)) power spectrum covariance matrix , 3 (right)) bispectrum covariance matrix, 4) power spectrum and joint power spectrum-bispectrum cumulative S/N. For this analysis we keep our usual binning in $\log \ell$ (14 regularly spaced bins from $\ell=10$ to $\ell=5000$) and all the survey related parameter as chosen in Appendix~\ref{App:EuclidSpec}. Specifically, we use a 5 bin tomography for the sources.}
\label{fig:Scatter}
\end{figure}

\subsection{Joint signal-to-noise ratio}
In Fig.~\ref{fig:SNFULLdep}, we finally move to the joint study of the probes. A joint analysis allows us to improve the forecast by accessing the cosmological information that was lost in mode couplings. Motivated by our previous findings, we restrict the joint analysis to 5 tomographic bins. In the case of the joint power spectrum-bispectrum S/N our data vector has dimension 10000, 818 and 144 when considering 5, 2 or 1 tomographic bins respectively. Comparing the joint S/N, NGssc included (black solid line), with the power spectrum alone case, no approximations assumed (blue solid line), we find that the maximum information content increases by $\sim $10\% for all the considered tomographic analysis. This confirms the need for the inclusion of the bispectrum analysis for future weak lensing analyses. 
Similarly to the previous paragraph, we address the effect of the super-sample covariance: when included, the maximum achievable information content is reduced (compared to the yellow solid line) by about $30\%$ in the 1 bin tomographic case and by about $25\%$ in the 2 and 5 bin analyses.
As a final remark on this joint analysis, we compare the information content in all the 3 tomographic cases with the hypothetical case of a Gaussian field contaminated with shot noise (dark green solid lines). If the convergence field were Gaussian, this line would represent a perfect reconstruction of the cosmological information in the field. It can be observed that, in the most informative case with 5 tomographic bins, we recover about 60\% of this ideal cosmological information.

\subsection{Uncertainty of the theoretical model: scatter of the halo concentration parameter}
We investigate the robustness of our S/N-based forecasts against the uncertainties of the halo profile properties. In particular, we want to test the impact of the convolution of the halos with the function $p\left(c_{\mathrm{v}},m,z\right)$ \eqref{LogNormC} which gives the probability that a halo of mass $m$ has a concentration parameters $c_{\mathrm{v}}$. Specifically it allows to account for the scatter of halos, as observed in simulations, around the mean relation $c_{\mathrm{v}}-m$ when fitted with an NFW profile ~\citep{2000ApJ...535...30J,2001MNRAS.321..559B}. In this section we make the redshift dependence explicit again. We compare the observables as built from two models. In the first model (model A), we evaluate $I_{\mu}^{\beta}$~\eqref{ImunuHalos} from the full integration over $c_\mathrm{v}$. In the second model (model B) we simply assume $p\left(c_\mathrm{v},m,z\right) = \delta_{\mathrm{D}}\left(c_\mathrm{v}-\bar{c}_{\mathrm{v}}\left(m,z\right)\right)$. This last approximation is the one mostly used in the literature. The median value $\bar{c}_{\mathrm{v}}\left(m,z\right)$ is given by the following fitting formula \citep{2001MNRAS.321..559B}
\begin{align}
        &\bar{c}_{\mathrm{v}}\left(m,z\right) = K\ \frac{a(z)}{a_{\mathrm{c}}}, \\
        &\hspace{3cm}m_{\star}\left(a_{\mathrm{c}}\right) = F\ m,\quad \nu\left( m_{\star},a_{\mathrm{c}}\right) = 1
\end{align}
where $a_c$ is the epoch at which the typical collapsing mass $m_{\star}$ (defined by $\nu=1$) equals a fixed fraction of the halo mass $m$ at the same epoch. The best-fit values for the free parameters $F,\ K$ are respectively 0.01 and 4.0 for a $\Lambda$CDM Cosmology.
In Fig.~\ref{fig:Scatter} (starting from the top), we can see that the vector of power spectra $\vec{P}$ differs at maximum by 1\% between the two models (first panel). The data vector $\vec{B}$ is more affected with a peak at 5\% (second panel) for the smallest scale. We recall from the introduction to this Section that the observables in our data vector are ordered for increasing value of the multipoles used within the configurations. The jagged profile instead corresponds to different tomographic dispositions of the sources per fixed spatial configuration. 
In the third row we compare the power spectrum (left panel) and the bispectrum (right panel) covariance matrix as derived from the two models. They differ at maximum by 4\% and 25\%, respectively. However, there is no impact at all at the level of reconstructed S/N: we can see in the bottom panel in Fig.~\ref{fig:Scatter} that the fractional differences between the two models is well within the numerical precision of our pipeline ($\sim$1\%).

For completeness, we refer to \cite{2001ApJ...554...56C} for a similar analysis. In this work, the impact of marginalising over the concentration parameter was performed at the level of three-dimensional, 1-halo power spectrum and trispectrum. Specifically they found an impact of $\sim$~5\% and $\sim$ 20\% respectively for a distribution $p\left(c_{\mathrm{v}},m,z\right)$ of variance $\sigma_{\ln c_{\mathrm{v}}} = 0.2$. 

\begin{figure}
\centering
\includegraphics[width=\hsize,clip=true]{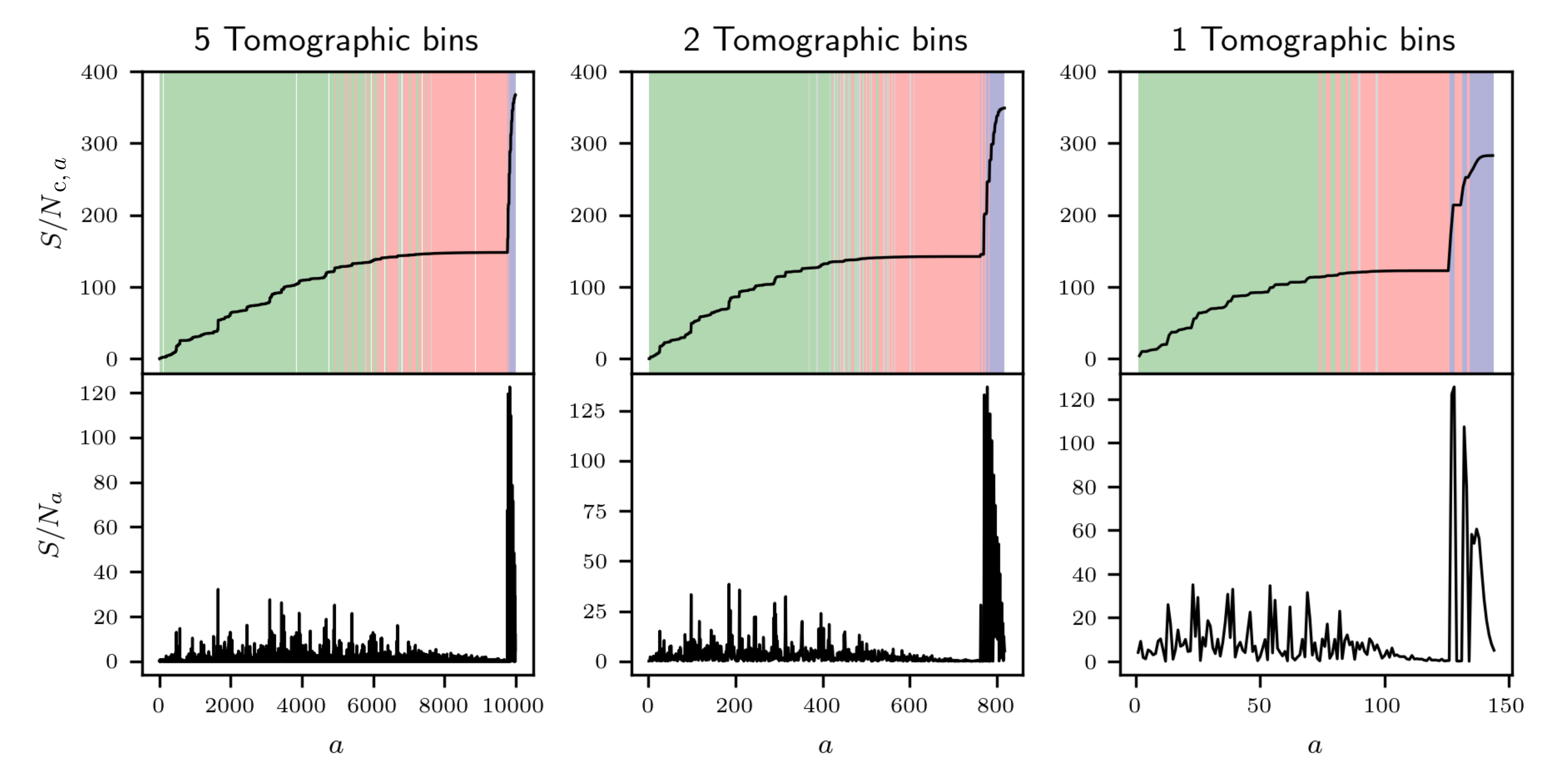}
\caption[]{PCA analysis for the joint 5 tomographic bin covariance matrix with no approximations on the covariance nor on the bispectrum data vector. \textit{Top:} $S/N_{\mathrm{c}, a}$ \eqref{CumulSN} as cumulative function of the eigenmodes $\vec{v}_a$. The eigenmodes $\vec{v}_a$ are ordered for increasing value of the associated variance $\lambda_a$.  The different colours on the background label modes which corresponds to observables $\hat{D}_a$ which are mostly combinations of power spectra (\textit{blue}), bispectra in the linear/mildly non-linear regime $[\ell < 400]$ (\textit{red}) and bispectra in the non-linear regime $[\ell > 400]$ (\textit{green}). We perform this classification for a given observable $\hat{D}_a$ by looking at which are the most representative observables $D_i$ contributing via the matrix $S_{ai}$. Considering only observables $D_i$ for which the corresponding matrix element $|S_{ai}| \geq 0.05$, the mode $\vec{a}$ is then classified according to which of the above classes represents more than the 90\% of them. If there is not a specific preference, we assign the \textit{grey} colour.  \textit{Bottom:} S/N ratio per eigenmode.  The eigenmodes $\vec{v}_a$ are ordered for increasing value of the associated variance $\lambda_a$.} 
\label{fig:PCAredcov1}
\end{figure}
\begin{figure}
\centering
\includegraphics[width=\hsize,clip=true]{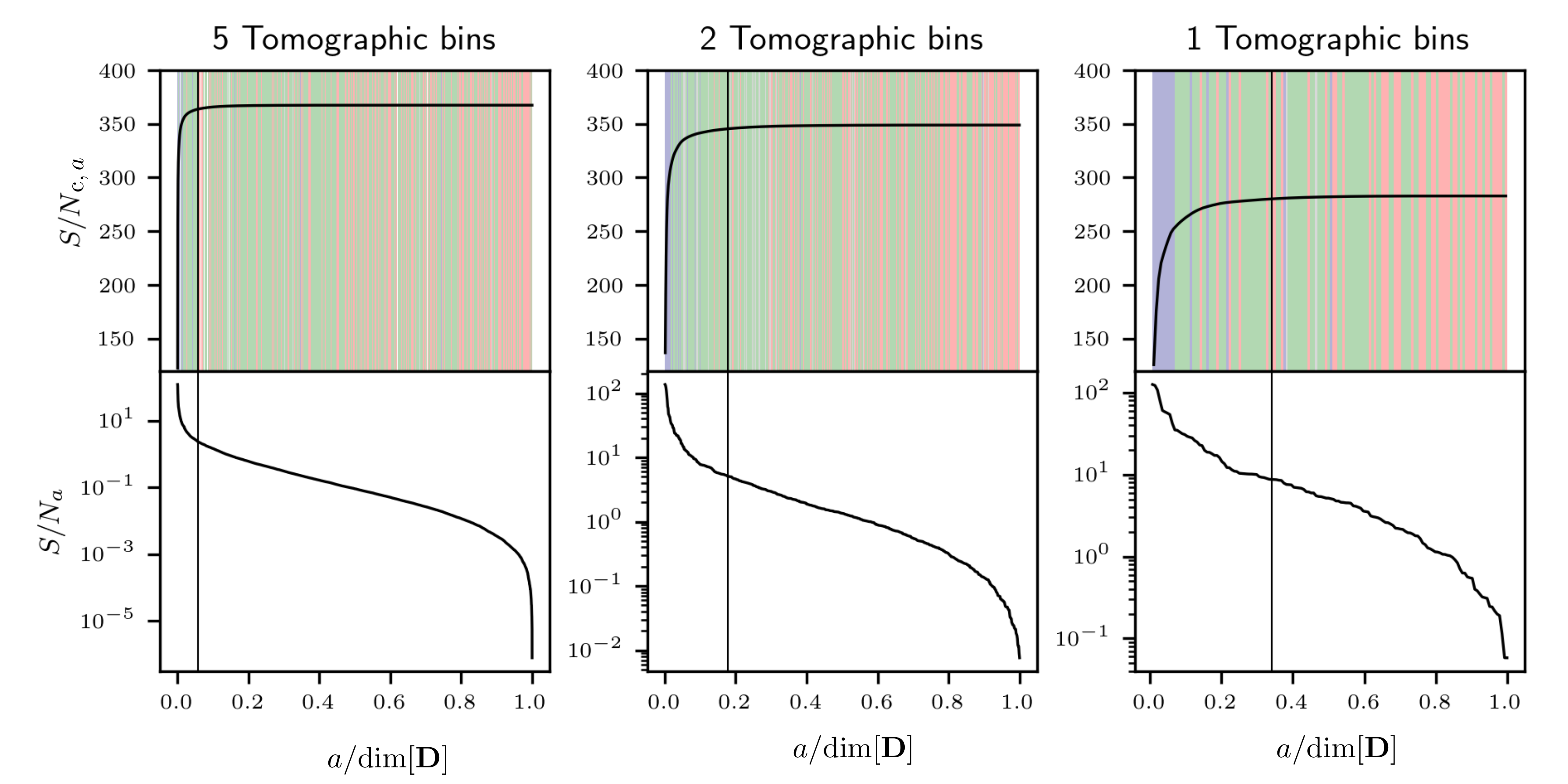}
\caption[]{The analysis here proposed is the same as in Fig.~\ref{fig:PCAredcov1}. However, the eigenmodes are now ordered for decreasing value of the associated $S/N_a$ and we are reporting our results in terms of the fraction of the total number of eigenmodes used. This is just motivated by graphical reasons. \textit{Top:} S/N as cumulative function of the eigenmodes~\eqref{CumulSN}. The colour code is the same as in Fig.~\ref{fig:PCAredcov1}. \textit{Bottom:} $S/N_a$  per eigenmode~\eqref{NewObs}. The vertical lines in the different panels indicate the fraction of eigenmodes required for recovering the 99\% of the maximum achievable S/N.} 
\label{fig:PCAredcov2}
\end{figure}
\begin{figure}
\centering
\includegraphics[width=\hsize,clip=true]{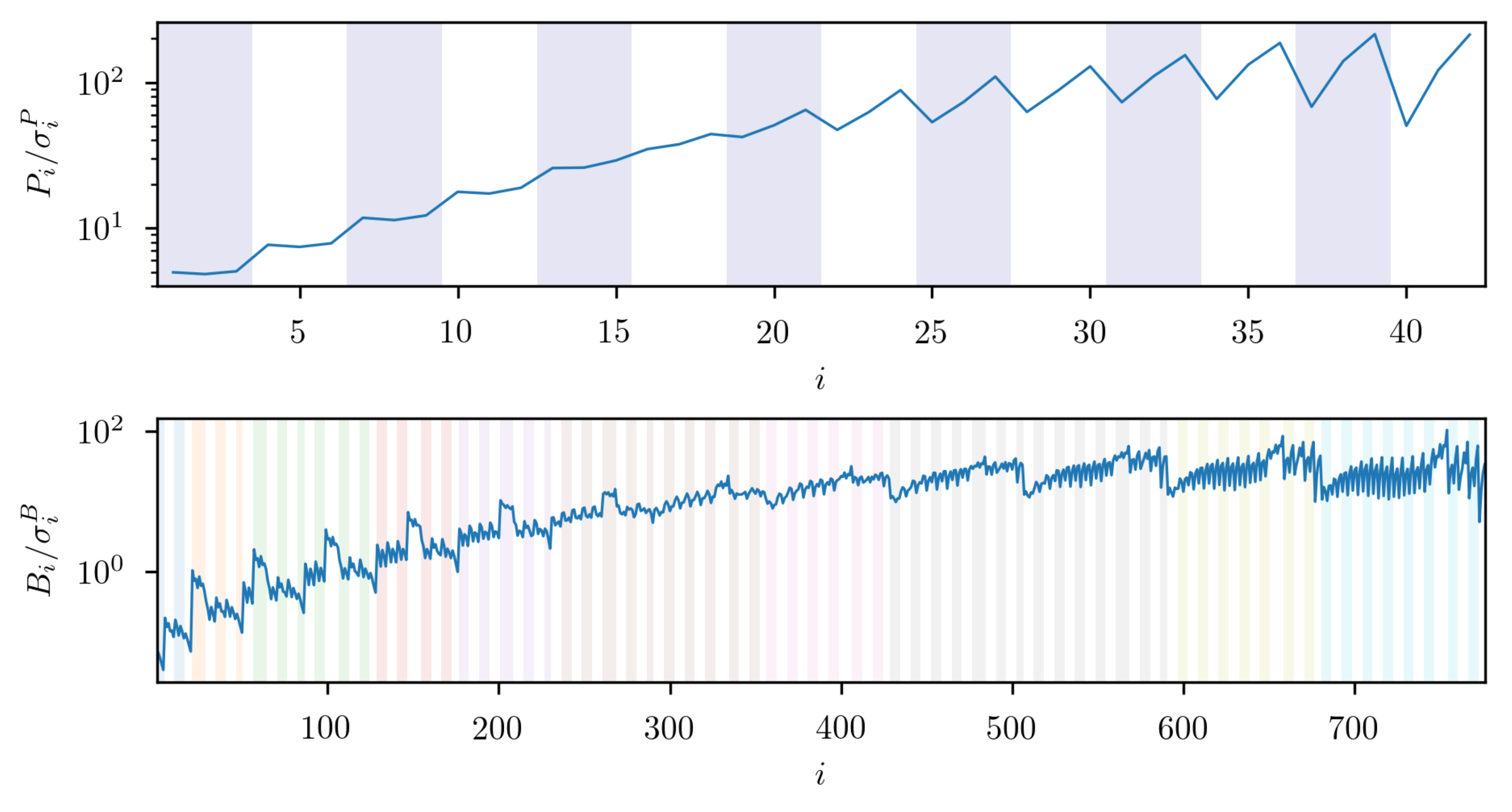}
\caption[]{Classical signal-to-noise ratio analysis for the original data vector $\vec{D}$. We diplay the ratio between the value of an observable, distinguishing between power spectra and bispectra, and the associated standard deviation  $\sigma^{P(B)}_i\equiv\sqrt{C^{PP(BB)}_{ii}}$. We assumed a 2 bin tomography. For every configuration in Fourier space, the different tomographic distributions of the sources are indicated by the same background colour. \textit{Top}: binned tomographic power spectrum. \textit{Bottom}: binned tomographic bispectrum. In the latter case, different colours on the background correspond to triangular configurations sharing the largest edge. Therefore, according to the ordering of the data vector, within each of these bends we move from squeezed to equilateral configurations.}
\label{fig:SN_diag}
\end{figure}

\section{Principal component analysis}\label{sec:PCA}
The bispectrum only brings a relatively small improvement of information, mainly due to the effect of the super-sample covariance. Thus, it seems desirable to simplify our problem and restrict our data vector to the most informative modes. The principal component analysis (PCA) is a simple way to exhibit which observables, or linear combinations of them, are the most informative.
It can be achieved via an eigenvalue decomposition of the $n\times n$ covariance $C$ onto an orthonormal basis of vectors $\vec{v}_{i=1,\dots,n}$
\begin{equation}
\label{SNPCA}
    C_{ij} = \sum_a\ S_{ai}\ S_{aj}\ \lambda_a.
\end{equation}
In the above equation, $S_{aj} \equiv \mathrm{v}_{a,j}$ and $\lambda_a$ is the $a^{\text{th}}$ eigenvalue associated to the mode $\vec{v}_a$. Also, we are omitting the supescrit D: from now on we apply this method only to the joint covariance implying $C\equiv C^{D}$. The decomposition \eqref{SNPCA} allows us to find linear combinations of the original observables in the data vector $\{D_i\}_{i=1,\dots,n}$ which define a new set of uncorrelated observables $\{\hat{D}_i\}_{i=1,\dots,n}$
\begin{equation}
\label{NewObs}
    \hat{D}_i = \sum_j\ S_{ij}\ D_j.
\end{equation}
At this point, the signal-to-noise ratio for each of them is simply the ratio between the signal and its own error
\begin{equation}
    \label{SNperEign}
    \left(\frac{S}{N}\right)^2_{a} = \frac{\hat{D}_a^2}{\lambda_a}.
\end{equation}
In the PCA formalism, the eigenvalue $\lambda_a$ is indeed the variance associated to $\hat{D}_a$.
From a much more interesting perspective, we can analyse the information content in the different modes $\vec{v}_a$ of the covariance as cumulative function of the new observables 
\begin{equation}
\label{CumulSN}
 \left(\frac{S}{N}\right)^2_{\mathrm{c}, a_{\mathrm{max}}} \equiv \sum_{a=1}^{a_{\mathrm{max}}} \left(\frac{S}{N}\right)^2_{a},
\end{equation}
up to a mode $\vec{v}_{a_{\mathrm{max}}}$. In the following, we analyse the cosmological information both as quantified from a single mode (Eq.~\eqref{SNperEign}) and as cumulative function of a set of eigenmodes (Eq.~\eqref{CumulSN}). In this kind of analyses, the smallest eigenvalues (associated to the most important eigenmodes in terms of information reconstruction) are potentially affected by  numerical errors. Due to the wide dynamics of the observables used for this work, the covariance matrices have large condition numbers: the eigenvalues span over a range of about $30$ orders of magnitude and the smallest ones can be affected by numerical errors if a too naive eigenmode decomposition algorithm is used. For this reason we rely on a specific high performance routine able to search for the eigenvalues (and associated eigenmodes) in a large dynamical range\footnote{Specifically we made use of Intel MKL Extended Eigensolver \texttt{dfeast\_syev} and implemented a search of the desired eigenvalues for each order of magnitude.}.

We present the main results of the PCA analyses in Fig.~\ref{fig:PCAredcov1} and in Fig.~\ref{fig:PCAredcov2} where the eigenmodes $\vec{v}_a$ are respectively ordered for increasing value of their variance $\lambda_a$ and decreasing information content $S/N_a$. The analyses have been applied on the full joint covariance, including super-sample contributions and all the multi-halo configurations for the bispectrum vector. In the first row of both figures, we display the reconstruction of the signal-to-noise ratio as cumulative function of the eigenmodes included in the covariance. In the second row of both figures instead, we show the information content per eigenmode $S/N_a$. We refer to the corresponding captions for more details. 
From a parallel study of Fig.~\ref{fig:PCAredcov1} and Fig.~\ref{fig:PCAredcov2} we can investigate the possibility of further reducing the dimensionality of our analysis while preserving the maximum information content. These figures ought to be analysed in the light of Fig.~\ref{fig:SN_diag} where we show the ratio between the (original) observables within the data vector $\vec{D}$ and their standard deviations $\sigma^{D}_i\equiv\sqrt{C^{D}_{ii}}$ (for a 2 bin tomography, for simplicity). In particular, Fig.~\ref{fig:SN_diag} would represent the information content of our observables if they were not correlated. Even though it does not provide a reliable insight on the strength of our measurements in the  regime we are exploring,  it can still help in understanding what are the configurations expected to be more (less) informative. Also, it helps in giving a physical understanding to the results of the PCA analyses.

First of all, by looking at the bottom row in  Fig.~\ref{fig:PCAredcov1}, we can identify a fraction of the eigenmodes which have a very poor information content $S/N_a$. On the top row, they correspond to a red plateau, i.e. eigenmodes mainly associated to bispectra (in the original data vector $\vec{D}$) in the linear/mildly non-linear regime via the linear combination \eqref{NewObs} (see caption for more details). This feature is present for all the tomographic redistributions of the sources. This result can be easily explained by looking at Fig.~\ref{fig:SN_diag}. As a matter of fact, the bispectra sourcing these modes have a very a low signal-to-noise ratio since at these scales the matter field has a very weak deviation from the Gaussian statistics. 

\begin{figure}
\centering
\includegraphics[width=\hsize,clip=true]{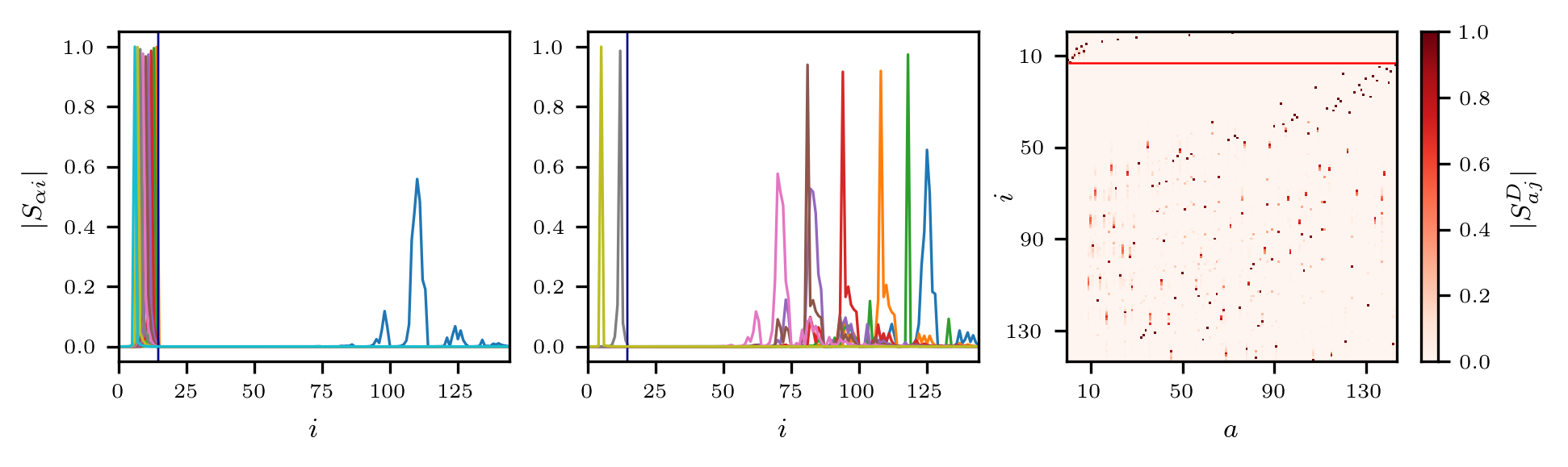}
\caption[]{\textit{Left}: values of the projection matrix $|S_{ai}|$ for different choices of $a$. For this plot, we picked the most informative modes $\mathbf{v}_a$ which allow to recover the 90\% of the total information content. We depict our results for the 1 tomographic bin joint covariance matrix. The vertical blue lines separates the vectors $\vec{P}$ (on the left) and $\vec{B}$ (on the right) within the vector of observables, labelled by $i$. \textit{Center}: same analysis as in the left panel. Here we picked the most informative modes $\mathbf{v}_a$ that allows us to recover an extra 5\% of the total information on top of the modes displayed in the left panel. \textit{Right}: representation of the full projection matrix  $|S_{ai}|$, $a$ being ordered for decreasing value of the information content per mode $S/N_a$. The red horizontal line separate power spectra and bispectra in the data vector.}
\label{fig:Shapes1}
\end{figure}
\begin{figure}
\centering
\includegraphics[width=\hsize,clip=true]{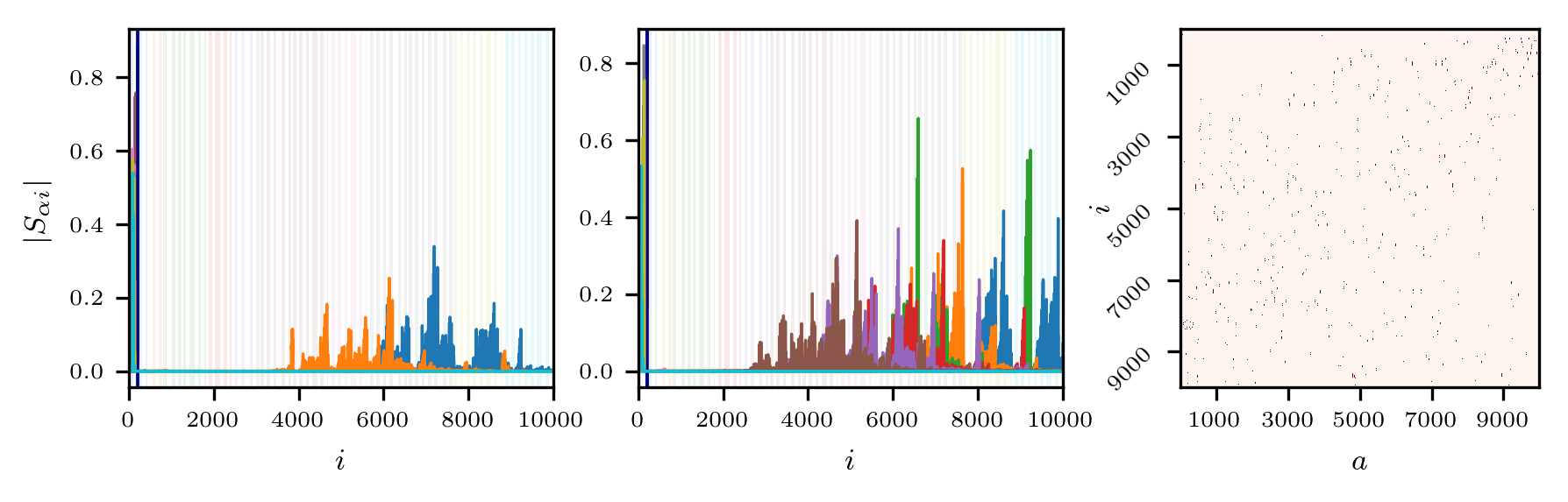}
\caption[]{\textit{Left}-\textit{Center}: same analysis as in the left-central panels in Fig.~\ref{fig:Shapes1} but for the 5 tomographic bin covariance matrix. The tiny different bends on the background include different tomographic contributions for the same Fourier configuration. Different colours on the background correspond to triangular configuration sharing the largest multipole. Therefore, according to the ordering of the data vector $\vec{D}$, within each of these bends we move from squeezed to equilateral configurations (more detailes within the main text).  \textit{Right}: sparsity pattern for the full projection matrix  $|S_{ai}|$ with $a$ being ordered for decreasing value of the information content per mode $S/N_a$. The points have been selected whenever $|S_{ai}| \geq 0.05$.}
\label{fig:Shapes5}
\end{figure}

Going back to the top-row panels of Fig.~\ref{fig:PCAredcov1}, the eigenmodes located in blue regions are characterised by the largest variance, being on the right end of all the panels. However, they are also the most important in terms of recovering the total information content, carrying the largest signal-to-noise ratio per mode (bottom row). In particular, they are at the left end of the panels in Fig.~\ref{fig:PCAredcov2}. Following the procedure described in detail within the caption of Fig.~\ref{fig:PCAredcov1}, we find that these modes are mainly linear combinations of power spectra. Then, we can easily understand the location of these modes within both Fig.~\ref{fig:PCAredcov1} and Fig.~\ref{fig:PCAredcov2}. Power spectra have an absolute standard deviation much larger than the bispectra ($\sim 10^{-9} - 10^{-13}$ and $\sim 10^{-12} - 10^{-20}$ respectively), which motivates their location on the right end of Fig.~\ref{fig:PCAredcov1} (we recall that, approximately, $C^{PP}\sim P^2$). On the other hand, power spectra have a much larger signal-to-noise ratio when compared to all the other observables in the vector $\vec{D}$. By comparing the first and the second row in Fig.~\ref{fig:SN_diag} we clearly see that power spectra can be measured with a much smaller statistical uncertainty when compared to the bispectra.

Finally, the modes located in the green bands are mainly sourced by bispectra evaluated on modes deep into the non-linear regime (see caption to Fig.~\ref{fig:PCAredcov1}). Compared to the modes associated to the bispectra in the linear/mildly non-linear regime (red), they have a lower variance, as they are located at the left end of the plots in Fig.~\ref{fig:PCAredcov1} while carrying more information: they represents a transition between the blue and the red modes in the top panels of Fig.~\ref{fig:PCAredcov2} and they are crucial for improving the information carried by the first modes associated to the power spectra (blue). Once again, we can understand these dynamics with the help of Fig.~\ref{fig:SN_diag}, where the bispectrum at non-linear scales has a higher signal-to-noise ratio compared to the one at larger scales. On the other hand, the smaller variance of these configurations is due to the fact that the bispectrum signal is much weaker the more we measure it in the non-linear regime, as we can see in Fig.~\ref{fig:spectra} (we recall that, approximately, $C^{BB}\sim P^3 + B^2$).

The vertical lines in Fig.~\ref{fig:PCAredcov2} indicate the fraction of eigenmodes required to recover 99\% of the full information content, once the PCA-modes have been reordered by decreasing value of the associated signal-to-noise ratio. The result is remarkable: the higher is the number of tomographic bins, the higher is the compression efficiency. Specifically, just $\lesssim$ 40\%, $\lesssim$ 20\% and $\lesssim$ 10\% of the modes are required  respectively for the 1, 2 and 5 bin analyses.

\begin{figure}
\centering
\includegraphics[width=\hsize,clip=true]{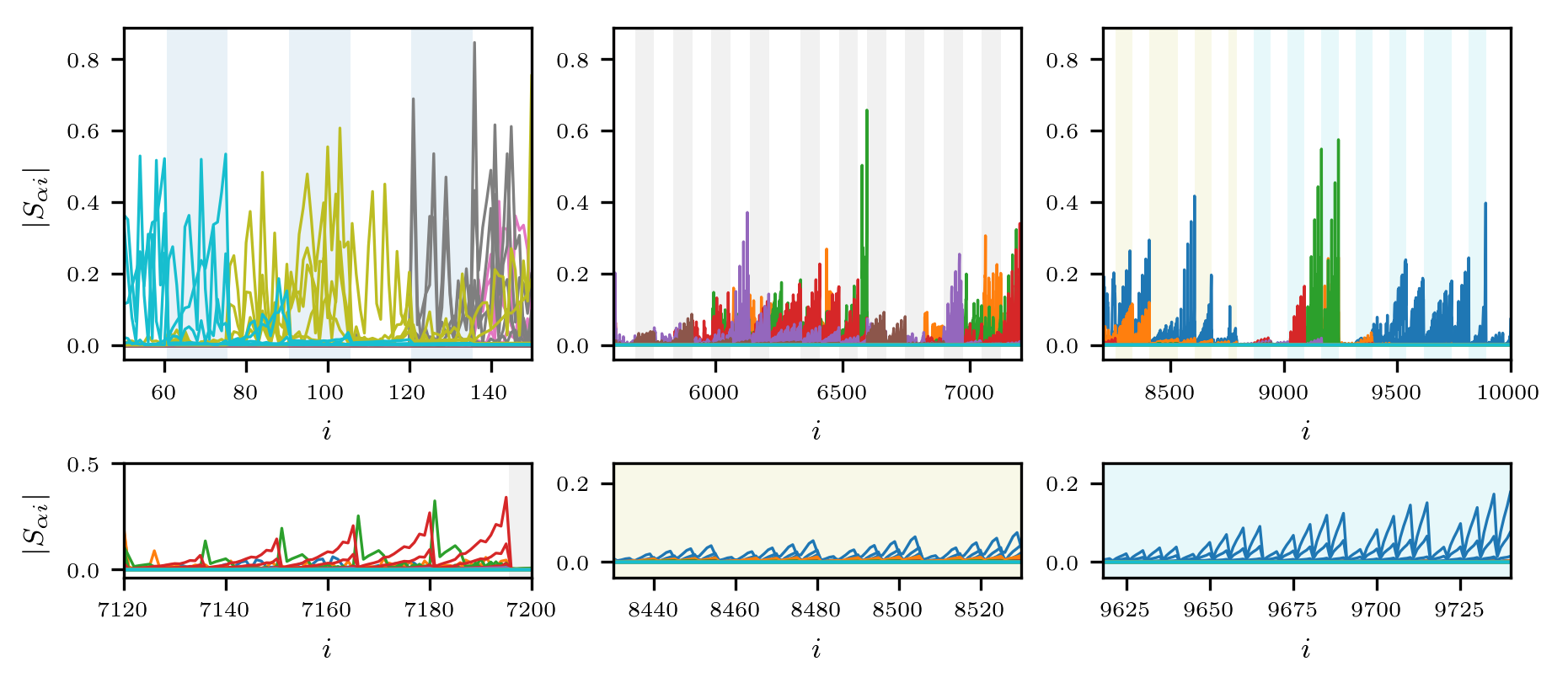}
\caption[]{Same analysis as in the left and central panel in Fig.~\ref{fig:Shapes5}. In particular we are considering a 5 tomographic bin joint covariance matrix. In the bottom panels, we are zooming on just one triangular configuration per panel. The colour map used in these plots is the same as in the left and central panels in Fig.~\ref{fig:Shapes5}, to facilitate the comparison.}
\label{fig:Shapes5_tomo}
\end{figure}
\begin{figure}
\centering
\includegraphics[width=\hsize,clip=true]{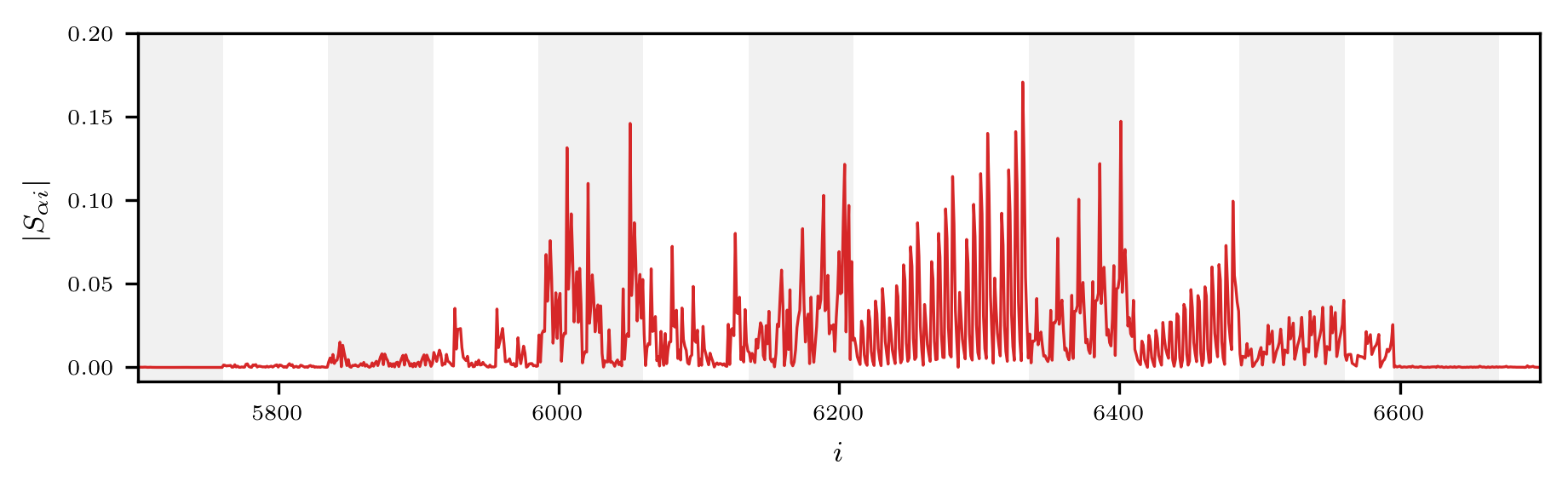}
\caption[]{Profile of the projection matrix $|S_{ai}|$ for a specific choice of mode $\vec{v}_a$ picked from the most informative ones accounting for the  95\% of the total information content. The colour code is the same as in Fig.~\ref{fig:Shapes5} in order to facilitate the comparison.}
\label{fig:tomo5_1mode}
\end{figure}

In Fig.~\ref{fig:Shapes1} and in Fig.~\ref{fig:Shapes5} we visualise the absolute value of the elements within the projection matrix $S_{ai}$~\eqref{SNPCA} for different choices of $a$ (left and central panel) and the total matrix itself (right panel). The index $a$ indicate a specific mode $\vec{v}_a$ of the original covariance matrix. In the right panels, the modes are ordered for decreasing value of the information content $S/N_a$. The index $i$ runs instead over the different elements of the data vector $\vec{D}$. Therefore, we want here to visualise how much each of the original observables contributes to the newly defined ones $\hat{\vec{D}}$. The colours on the background of the left and central panels have to be read as follows. Different tiny bends (alternatively white and coloured) refer to the same configurations in Fourier space, then spanning over the different tomographic contributions to it. We recall that in our data vector $\vec{D}$ these contributions are contiguous. In the region of the x-axis corresponding to the sub-vector $\vec{B}$ (on the right of the vertical violet line), different colours of the bends indicate Fourier configurations sharing the same larger multipole. In particular, while moving from left to the right within each of these macro-bends, we are actually spanning bispectrum configurations from squeezed triangles to equilateral ones.
In Fig.~\ref{fig:Shapes1}, we first focus on the 1 tomographic bin joint covariance matrix. In the left panel we depict the values $|S_{ai}|$ as function of the index $i$ for the most informative modes $\vec{v}_a$ recovering the 90\% of the total information content. We can see that they are mainly sourced by the power spectra in the non-linear regime. In the central panel we show the values $|S_{ai}|$ as function of the index $i$ for the modes $\vec{v}_a$ required to add an extra 5\% on top of the previous ones. They are mainly linear combinations of bispectra in the non-linear regime. Furthermore, the projection matrix shown in the rightmost panel is extremely sparse, peaking on just few configurations: overall, just 53 observables (36\%) are involved in recovering 95\% of the information content\footnote{\label{note1}We consider configurations $D_i$ with an associated weight $|S_{ai}| < 0.05$ as negligible. The same threshold has been applied to reproduce the sparsity pattern for the 5 tomographic bin covariance matrix in Fig.~\ref{fig:Shapes5}.}. This open a possibility for a second level of data compression beyond the one identified at the previous step at the level of modes.
In Fig.~\ref{fig:Shapes5} we perform the same analysis for the 5 bin tomography, reaching similar conclusions. In particular the tomography does not affect the possibility of dimensional reduction: we can discard 2093 observables from the original data vector (i.e. 80\% of it) and still recover 95\% of the information\footnoteref{note1}. 

In Fig.~\ref{fig:Shapes5_tomo}, we show the profiles $|S_{ai}|$ zooming onto smaller fraction of the data vector $\vec{D}$ (top row) and onto single triangular configurations (bottom row). We are here considering the 5 bin tomography covariance. The first panel on the left in the top row focus on the region of x-axis associated to the vector $\vec{P}$. These zoomed-in panels are useful in order to understand the distribution of the cosmological information among different tomographic bins, per fixed spatial configuration. While it is not possible to identify dominant tomographic configurations at the level of the power spectra (top-leftmost panel), at the bispectrum level the tomographic configurations that are most informative are, in general, those involving high redshift sources. We can clearly see this feature in the bottom row of Fig.~\ref{fig:Shapes5_tomo} where we focus on single spatial configurations. The jagged profile mirrors the ordering of the tomographic dsitributions of the sources within the data vector. We recall that for the bispectrum, the observables associated to the tomographic bins $\left(i,j,k\right)$ are ordered such that $k$ is the fastest index and $i$ the slowest index. Then we can see a peak in the profile whenever, for a fixed value of $k>i,j$, $i$ and $j$ move closer to $k$. The maximum information for a given spatial configuration happens to be when all the tomographic indices have reached the maximum value corresponding to the furthest sources.

Finally, in Fig.~\ref{fig:tomo5_1mode}, we show the profile of the projection matrix $|S_{ai}|$ for a single mode $a$, picked among the most informative ones recovering the 95\% of the total information content in a 5 bin tomography. In particular we picked a mode mainly dominated by bispectra to inspect the possibility of reducing the computational burden for this part of the data vector. Unfortunately, we can see that there are no dominant configurations and bispectra evaluated on several different spatial configurations are required to access the information content of the mode analysed. Also, for a specific triangular configuration (single background bend), it is not possible to identify a dominant contribution from a specific distribution of the sources.
\section{Discussions and conclusions}
\label{sec:Conclusions}
As part of this final section, we summarise the approximations adopted throughout our analysis along with their range of validity. As we will explain, the precision of the halo model in describing the matter clustering can represent a bottleneck for the accuracy of the whole forecast. Therefore, while lacking a better semi-analytical model for the gravitational collapse in the non-linear regime, it is important to remember that the halo model itself is an approximation. 
To start with, in deriving the statistical properties of the projected fields we assumed the Limber and the flat-sky approximations. \cite{2017MNRAS.472.2126K} thoroughly explored these approximations at the level of 2-point statistics and compared their predictions against a full-sky approach for the cosmic shear field with a CFHTLenS-like galaxy distribution \citep{2013MNRAS.430.2200K}. For the joint approximation under exam, they found it to be accurate to better than 10\% for $\ell > 3$, converging slowly to the true projection with percent level precision at $\ell > 100$. Given that the accuracy of the halo model predicted polyspectra is in general much lower, we consider the above performance satisfactory for our work. Secondly, while computing the covariance matrix for the power spectrum and bispectrum we assumed that the trispectra, 5- and 6-order polyspectra in Eqs.~\eqref{CovPPNG},~\eqref{NeedThis3} and ~\eqref{NeedThis2} (resp.) are slowly varying within the considered $\ell$-bins. To our knowledge, no tests in the literature were performed on this matter for the observables and the scales of interests. However, given the dynamics of the spectra in Fig.~\ref{fig:spectra} (left panel) we advise further tests of this assumption, especially for high $\ell$. On the other hand, the high number of configurations required for our analyses makes the full integrations over the bin width unfeasible given the computational resources at our disposal. Still at the level of covariance matrix we approximated all the polyspectra with their 1-halo component (starting from the 3-point one). Once again, we are not aware of any studies on the impact of neglecting these higher halo terms. Given our discussion at the beginning of Sec.~\ref{sec:halomodel}, we do not expect them to be important for cosmological analyses even though they may have a role on regularising the covariance matrix reducing numerical errors at the inversion. Finally, in the computation of the super-sample covariance~\eqref{WLSSCflatskyLimber}, we are again assuming the flat-sky and  Limber approximations for both the intra- and super-survey modes. At the power spectrum level, \cite{2018JCAP...06..015B} analysed the impact of these approximations against a full spherical analysis of the super-survey modes. For Euclid- and LSST-like tomographies and at the power spectrum covariance level, they report that for surveys covering less than $\sim$ 5\% of the sky the two results agree to better than 1\%. However, when moving to the expected realistic coverage of $f_{\text{sky}} \approx 0.3-0.4 $ the use of the flat-sky expressions results in an underestimation of the SSC contribution of about 10\%. Again, this performance has to be considered satisfactory for the present work given that the precision bottleneck is mainly given by the accuracy of the matter clustering model.

For completeness, we also want to compare the analyses performed in this paper with similar works in the literature. The impact of the correlations between observed and super-sample modes has already been addressed in the literature. Specifically, \cite{2013arXiv1306.4684K} performed a preliminary study on the information content of the weak lensing bispectrum, which was further developed in \cite{2013MNRAS.429..344K} where 3 tomographic bins where considered for cosmological parameter forecast. However, they included the super-sample correlations at very small scales. These terms are known in the literature as halo sample variance and they contribute to the correlations when all the points in a given configuration are inside the same halo. Formally, the halo sample variance is just a part of the components we obtained via the response approach used for our implementation. In particular they corresponds to the last terms in Eq.~\eqref{PowerHMResp} and Eq.~\eqref{BispHMResp} which are the responses of the 1-halo term to the long mode $\delta_{\mathrm{b}}$. 
In a different study, \cite{2019JCAP...03..008B} accounts for the super-sample covariance in the PT response formalism by making different approximations. In particular, he considers only correlations  between bispectra on squeezed configurations. This approximation results in a negligible impact on the signal-to-noise ratio when adding the super-sample covariance. Our study shows that this is not the case  when all the triangular configurations are included in the analysis.

Finally, we summarise the results of our work. We have presented the first full joint analysis of the information content for the convergence weak lensing power spectrum and bispectrum for a Euclid-like survey. We modelled the observables with the halo model and developed a high performance code capable to output fast and precise covariance matrices for the binned tomographic spectra. The covariance matrices calculated for the present work have been used for signal-to-noise ratio analyses, but can benefit any forecast based on a Gaussian likelihood. 

At the power spectrum level and including shot noise, we were capable to show that the super-sample covariance is the main source of error, leading to a reduction of 40\% of the maximum achievable signal-to-noise ratio compared to the Gaussian case. The non-Gaussian cross-correlations between in-survey modes  account for a loss of information of about 30\% instead when compared to the Gaussian case. The combined effect of these two sources of error leads to a loss of about 45\% on the signal-to-noise ratio. 
An important result of our analysis is the possibility to recover the cosmological information content of a Euclid-like survey by using 5 equipopulated tomographic redshift bins, instead of 10. This result is in particular insensitive to the angular scale and to the components included in the error budget of the observables and does not account for systematics like PSF, photo-z, blending or intrinsic alignments. The same phenomenology was found at the bispectrum level. On the bispectrum signal-to-noise ratio the super-sample covariance has an impact of about $\sim$13\% (10 bin tomography) when the observables have been estimated via all the multi-halo configurations. We tested the impact of these configurations in the modelling of the vector of bispectra and we found that neglecting the 3-halo terms is a good approximation for our analysis, both when including or excluding the super-sample covariance.

Motivated by our previous findings, we performed a joint power spectrum-bispectrum analysis on 5 equipopulated tomographic redshift bins. We proved that this combined approach can improve the information content by $\sim$10\% with respect to the power spectrum alone. The super-sample covariance of the bispectrum cannot be ignored  and reduces the maximum information achievable of about $\sim$25\%. In these analyses the 2- and 3-halo terms have been taken into account for the bispectra in the vector, but we restricted the computation of the covariance to the 1-halo terms which dominates the covariance for the most relevant configurations. These are the most important results of this work.

We found that the halo modelling uncertainty due to the scatter of the concentration parameter in the simulations does not affect the joint signal-to-noise ratio by more than 1\%.

In this work, a preliminary study on the possibility of further compressing our covariance matrices while preserving the cosmological information is considered. In particular, by performing a principal component analysis on our covariance matrices, we found that a very small fraction of the eigenmodes ($\lesssim$ 10\% for 5 bin tomography) carries most of the information, and that not all the configurations equally contribute to the full information content. Indeed, only 20\% of the data vector is of importance for the linear combinations that form this 10\% of eigenmodes, and thus a large fraction of the vector of the observables ($\sim$ 80\% for 5 bin tomography) is not significant to reconstruct the signal-to-noise ratio at the different scales considered in this work.

Starting form the results presented in this paper, several possible research paths are open. The natural and most important step forward is a full cosmological parameter forecast in order to translate the analyses on the signal-to-noise ratio into actual error bars on the 
parameters. Secondly, the expressions used to model the super-sample correlations are based on the flat-sky and the Limber approximations. However, they might not be precise for the volumes accessible with future galaxy surveys and further tests are advised. Finally, the actual implementation of data compression techniques qualitatively identified in the last part of this work would lead to great benefit for future data analyses. In particular, with regards to the numerical implementation of future likelihoods, the simplification of covariance matrices for joint 2- and 3-point statistics should be consider as a high priority task to face in order to be ready for upcoming data.

\bibliography{refs}

\section*{Acknowledgements}
We would like to express our very great appreciation to Alexandre Barreira, whose crucial suggestions were of great help in identifying an important weakness in a first version of the manuscript.
We would also like to thank Sandrine Codis, Silvia Galli, Benjamin Joachimi and Susan Pyne for fruitful discussions and interesting comments on our results. We convey our gratitude to Doogesh Kodi Ramanah for constructive suggestions to improve the manuscript.
This work has made use of the Horizon Cluster hosted by Institut d'Astrophysique de Paris. We thank Stephane Rouberol for maintenance and running of this computing cluster. We made use of the public code \verb"HMcode" developed by ~\cite{2015MNRAS.454.1958M} and available at \url{https://github.com/alexander-mead/hmcode}. We acknowledge financial support from the ILP LABEX (under reference ANR-10-LABX-63) which is financed by French state funds managed by the ANR within the Investissements d'Avenir programme under reference ANR-11-IDEX-0004-02. In particular, M.R. acknowledges financial support from the Centre National d'Etudes Spatiales (CNES) fellowship program. F.L. acknowledges support by the Swiss National Science Foundation.








\appendix

\section{A Euclid-like survey}
\label{App:EuclidSpec}
While analytical results in this work apply to any weak lensing survey, we illustrated them with specific application to the Euclid mission. To this end, we make use of the requirements presented in \cite{2011arXiv1110.3193L}. Specifically, the angular multipole range we investigate is $\left[10,5000\right]$, and we chose to bin it in 14 regularly spaced intervals in $\log \ell$, then with extrema (approximately): 10, 16, 24, 38, 59, 92, 143, 224, 348, 543, 847, 1320, 2058, 3208, 5000. For the sky coverage, we use $\Theta_{\text{sky}} = 1.29\ \mathrm{rad},\ \Omega_{\text{sky}}= 15.000\ \mathrm{deg}^2 \approx 4.57\ \mathrm{sterad},\ f_{\text{sky}} = 0.36$. For the photometric properties of the survey, we use a total comoving number of observed sources of $n_{\text{tot}} = 30\ \text{gal}\ \text{arcmin}^{-2}$ from $z_{\text{min}}=0.001$ up to $z_{\text{max}}=2.500$. The distribution of the sources is $n\left(z\right) \propto \left(z/z_o\right)^2 \text{exp}[-\left(z/z_o\right)^{1.5}]$ where $z_o=0.9/\sqrt{2}$, and it is normalized over the observed range. The sources are then split in (up to) 10 equi-populated redshift bins with extrema: 0.001, 0.418, 0.560, 0.678, 0.789, 0.900, 1.019, 1.155, 1.324, 1.576, 2.500.

\section{Joint covariance matrix for the weak lensing convergence power spectrum and bispectrum}
\label{App:covariances}
To complement the discussion of Sec.~\ref{sec:CorrCov}, we give here in detail the Gaussian and non-Gaussian error contributions to the tomographic bispectrum signal, including the cross-covariance with the power spectrum \citep{2013arXiv1306.4684K}. While in Sec.~\ref{3.2} we followed a didactic approach by first introducing the estimator $\hat{P}_{ij}(\ell^{\mathrm{b}})$~\eqref{PSest} and in a second moment the masked estimator $\hat{P}^{W}_{ij}(\ell^{\mathrm{b}})$~\eqref{PSWest}, here we introduce the bispectrum covariance directly with the general approach of the masked fields. Therefore, borrowing the notation from Sec.~\ref{3.2}, we introduce the following binned bispectrum estimator
\begin{multline}
\label{BSWset}
    \hat{B}_{ijk}^W\left(\ell_1^{\mathrm{b}}, \ell_2^{\mathrm{b}}, \ell_3^{\mathrm{b}}\right) \equiv \frac{1}{\Omega_{\text{sky}}\  N_{\text{tri.}}\left(\ell_1^{\mathrm{b}}, \ell_2^{\mathrm{b}}, \ell_3^{\mathrm{b}}\right)}\int\frac{d^2\Bell_1^{''}}{\left( 2\pi\right)^2}\int\frac{d^2\Bell_2^{''}}{\left( 2\pi\right)^2}\int\frac{d^2\Bell_3^{''}}{\left( 2\pi\right)^2} \tilde{W}( \Bell_1^{''})\tilde{W}( \Bell_2^{''})\tilde{W}( \Bell_3^{''})\times\\
    \sum_{\Bell'_1,\Bell'_2,\Bell'_3}\kappa_{\Bell'_1-\Bell^{"}_1}^{(i)}\kappa_{\Bell'_2-\Bell^{"}_2}^{(j)}\kappa_{\Bell'_3-\Bell^{"}_3}^{(k)}\Delta^{(3)}_{\ell_1^{\mathrm{b}}, \ell_2^{\mathrm{b}}, \ell_3^{\mathrm{b}}} \left(\Bell'_1,\Bell'_2,\Bell'_3 \right).
\end{multline}
 The newly introduced selection function $\Delta^{(3)}_{\ell_1^{\mathrm{b}}, \ell_2^{\mathrm{b}}, \ell_3^{\mathrm{b}}} \left(\Bell'_1,\Bell'_2,\Bell'_3 \right)$ forces on the modes $\Bell'_1, \Bell'_2, \Bell'_3$ the constraint  $\Bell'_i \in \ell^{\mathrm{b}}_i$ for $i=1,2,3$ simultaneously and further requires that $\Bell'_1 + \Bell'_2 + \Bell'_3 = 0$. $N_{\text{tri.}}\left(\ell_1^{\mathrm{b}}, \ell_2^{\mathrm{b}}, \ell_3^{\mathrm{b}}\right)$ normalises the sum over the total number of independent triangular configurations built with the modes available within the combined bins  \citep{2008A&A...477...43J,2009A&A...508.1193J,2013MNRAS.429..344K}. 
\begin{align}
N_{\text{tri.}}\left(\ell_1^{\mathrm{b}}, \ell_2^{\mathrm{b}}, \ell_3^{\mathrm{b}}\right) &= \sum_{\Bell'_1,\Bell'_2,\Bell'_3}\Delta^{(3)}_{\ell_1^{\mathrm{b}}, \ell_2^{\mathrm{b}}, \ell_3^{\mathrm{b}}} \left(\Bell'_1,\Bell'_2,\Bell'_3 \right) \approx \frac{2\Omega_{\text{sky}}\ \ell_1\ \ell_2\ \ell_3\ \Delta\ell_1^{\mathrm{b}}\ \Delta\ell_2^{\mathrm{b}}\ \Delta\ell_{(3)}^{\mathrm{b}}}{\sqrt{2\ \ell_1^2\ \ell_2^3\ \ell_3^2 - \ell_1^4 - \ell_2^4 - \ell_3^4}}
\end{align} 
where the approximation assumed is $\Delta\ell_i^{\mathrm{b}} \gg \ell_f$.  The calculation starts from the basic definition of covariance of two estimators $\mathcal{O}_1, \mathcal{O}_2$
\begin{equation}
    \label{CovMat1}
    \text{Cov}\left[\mathcal{O}_1,\mathcal{O}_2\right] = \langle\mathcal{O}_1\mathcal{O}_2\rangle - \langle\mathcal{O}_1\rangle\langle\mathcal{O}_2\rangle. 
\end{equation}
By replacing the power spectrum and the bispectrum estimators~\eqref{PSWest}-\eqref{BSWset} (resp.) within the covariance definition, we can derive the full bispectrum covariance and power spectrum-bispectrum cross-covariance. Similarly to Eq.~\eqref{CovPP1}, we reduce the correlations between the windowed instances of the convergence field into their connected components. In particular, the $\text{Cov}[BB]$ (schematically) requires the decomposition of a 6-point correlation function into the sum of its irreducible components: (2-$\times$2-$\times$2-), (3-$\times$3-), (2-$\times$4-) and 6-order polyspectra (qualitatively). We label them in the following with the further subscript Gauss, BB, PT and 6P respectively.
The $\text{Cov}[PB]$ requires the decomposition of a 5-point correlation function into the sum of its 5-point connected component and into a serie of power spectrum-bispectrum products. Analogously, we will label them with the further subscript 5P and PB respectively.  At the level of matter power spectrum-bispectrum covariance, \cite{2019arXiv190101243B} proved that the window function does only impact the 6- and 5-order polyspectra related terms respectively in  $\text{Cov}[BB]$ and  $\text{Cov}[PB]$. Since we are working under the flat-sky approximation, the conclusion drawn from the work of \cite{2019arXiv190101243B} naturally extends to the weak lensing field (by moving from  three-dimensional Fourier integrals to two-dimensional ones). A detailed calculation leads to the following expressions for the intra-survey part (NGins) of the bispectrum covariance 
\citep{2013MNRAS.429..344K,2013arXiv1306.4684K}
\begin{align}
\text{Cov}\Big[\hat{B}_{ijk}^{\mathrm{W}}\big(\ell_1^{\mathrm{b}},\ell_2^{\mathrm{b}},\ell_3^{\mathrm{b}}\big),&\hat{B}_{i'j'k'}^{\mathrm{W}}\left(\ell^{'\mathrm{b}}_1,\ell^{'\mathrm{b}}_2,\ell^{'\mathrm{b}}_3\right) \Big]_{\text{Gauss}} 
\\
=\frac{\Omega_{\text{sky}}}{N_{\text{tri.}}\left(\ell_1^{\mathrm{b}},\ell_2^{\mathrm{b}},\ell_3^{\mathrm{b}}\right)}\Big[ &P_{(ii')}^{\text{s.}}\left(\ell_1\right)\delta^{\mathrm{K}}_{\ell_1\ell'_1} \Big\{ P_{(jj')}^{\text{s.}}\left(\ell_2\right)P_{(kk')}^{\text{s.}}\left(\ell_3\right)\delta^{\mathrm{K}}_{\ell_2\ell'_2}\delta^{\mathrm{K}}_{\ell_3\ell'_3} + P_{(jk')}^{\text{s.}}\left(\ell_2\right)P_{(kj')}^{\text{s.}}\left(\ell_3\right)\delta^{\mathrm{K}}_{\ell_2\ell'_3}\delta^{\mathrm{K}}_{\ell_3\ell'_2} \Big\}\ \nonumber\\
&+ \text{2 terms obtained from perm. of}\ \big( i'\leftrightarrow j', \ell'_1\leftrightarrow \ell'_2 \big) \nonumber\\
&+ \text{2 terms obtained from perm. of}\ \big( i'\leftrightarrow k', \ell'_1\leftrightarrow \ell'_3 \big) \Big],
\end{align}
\begin{align}
\text{Cov}\Big[\hat{B}_{ijk}^{\mathrm{W}}\big(\ell_1^{\mathrm{b}},\ell_2^{\mathrm{b}},\ell_3^{\mathrm{b}}\big),&\hat{B}_{i'j'k'}^{\mathrm{W}}\left(\ell^{'\mathrm{b}}_1,\ell^{'\mathrm{b}}_2,\ell^{'\mathrm{b}}_3\right) \Big]_{\text{NGins,BB+PT}} \nonumber\\
= \frac{2\pi}{\Omega_{\text{sky}}}\Big[  &\frac{1}{\ell_1\Delta\ell_1^{\mathrm{b}}}\left(
B_{i'jk}\left(\ell'_1,\ell_2,\ell_3\right)B_{ij'k'}\left(\ell_1,\ell'_2,\ell'_3\right)\delta^{\mathrm{K}}_{\ell_1\ell'_1}+ 
B_{j'jk}\left(\ell'_2,\ell_2,\ell_3\right)B_{i'ik'}\left(\ell'_1,\ell_1,\ell'_3\right)\delta^{\mathrm{K}}_{\ell_1\ell'_2}+
B_{k'jk}\left(\ell'_3,\ell_2,\ell_3\right)B_{i'j'i}\left(\ell'_1,\ell'_2,\ell_1\right)\delta^{\mathrm{K}}_{\ell_1\ell'_3}\right) \nonumber \\
&+ \text{3 terms obtained from perm. of}\ \big( i\leftrightarrow j, \ell_1\leftrightarrow \ell_2 \big) \nonumber\\
&+ \text{3 terms obtained from perm. of}\ \big( i\leftrightarrow k, \ell_1\leftrightarrow \ell_3 \big)\Big] \nonumber \\
+ \frac{2\pi}{\Omega_{\text{sky}}}\Big[  &\frac{1}{\ell_1\Delta \ell_1^{\mathrm{b}}}\left(
P_{ii'}^{\text{s.}}\left(\ell_1\right)T_{jkj'k'}\left(\ell_2,\ell_3,\ell'_2,\ell'_3\right)\delta^{\mathrm{K}}_{\ell_1\ell'_1}+ 
P_{ij'}^{\text{s.}}\left(\ell_1\right)T_{jki'k'}\left(\ell_2,\ell_3,\ell'_1,\ell'_3\right)\delta^{\mathrm{K}}_{\ell_1\ell'_2}+
P_{ik'}^{\text{s.}}\left(\ell_1\right)T_{jki'j'}\left(\ell_2,\ell_3,\ell'_1,\ell'_2\right)\delta^{\mathrm{K}}_{\ell_1\ell'_3}\right) \nonumber \\
&+ \text{3 terms obtained from perm. of}\ \big( i\leftrightarrow j, \ell_1\leftrightarrow \ell_2 \big) \nonumber\\
&+ \text{3 terms obtained from perm. of}\ \big( i\leftrightarrow k, \ell_1\leftrightarrow \ell_3 \big)\Big] \label{CovBB}
\end{align}
while the covariances between the binned tomographic power spectrum and bispectrum are
\begin{align}
\label{Covpb}
\text{Cov}\Big[\hat{P}^{\mathrm{W}}_{ij}\left(\ell^{\mathrm{b}}\right),\hat{B}^{\mathrm{W}}_{i'j'k'}&\left(\ell^{\mathrm{b}}_1,\ell^{\mathrm{b}}_2,\ell^{\mathrm{b}}_3\right) \Big]_{\text{NGins,PB}} = \nonumber\\
= \frac{2\pi}{\Omega_{\text{sky}}}\Big[&\frac{1}{\ell_1\Delta\ell_1^{\mathrm{b}}}\left(
P_{i'j}^{\text{s.}}\left(\ell\right)B_{ij'k'}\left(\ell,\ell_2,\ell_3\right)\delta^{\mathrm{K}}_{\ell\ell_1}+ 
P_{ii'}^{\text{s.}}\left(\ell_1\right)B_{jj'k'}\left(\ell,\ell_2,\ell_3\right)\delta^{\mathrm{K}}_{\ell\ell_1}\right)+ \nonumber \\
&+ \text{2 terms obtained from perm. of}\ \big( i'\leftrightarrow j', \ell_1\leftrightarrow \ell_2 \big)\	+\nonumber\\
&+ \text{2 terms obtained from perm. of}\ \big( i'\leftrightarrow k', \ell_1\leftrightarrow \ell_3 \big)\Big].
\end{align}
For the moment, we did not include the components related to the 6- and 5-order polyspectra. As anticipated, these terms are indeed affected by the window function of our survey (see \cite{2018PhRvD..97d3532C,2019arXiv190101243B} for a three-dimensional analysis) as 
\begin{multline}
\text{Cov}\Big[\hat{B}_{ijk}^{\mathrm{W}}\big(\ell^{\mathrm{b}}_1,\ell^{\mathrm{b}}_2,\ell^{\mathrm{b}}_3\big),\hat{B}^{\mathrm{W}}_{i'j'k'}(\ell^{'\mathrm{b}}_1,\ell^{'\mathrm{b}}_2,\ell^{'\mathrm{b}}_3) \Big]_{\text{NG,P6}} = \\
     \sum_{\bar{\Bell}_1,\bar{\Bell}_2,\bar{\Bell}_3}\sum_{\bar{\Bell}'_1,\bar{\Bell}'_2,\bar{\Bell}'_3}\int\frac{d^2\Bell^{''}}{\left( 2\pi\right)^2}|W(\Bell^{''})|^2
    P_{ijki'j'k'}\left(\bar{\Bell}_1,\bar{\Bell}_2,\bar{\Bell}_3 + \Bell^{''}
    ,\bar{\Bell}'_1,\bar{\Bell}'_2,\bar{\Bell}'_3 - \Bell^{''}\right)
    \frac{\Delta^{(3)}_{\ell_1^{\mathrm{b}}, \ell_2^{\mathrm{b}}, \ell_3^{\mathrm{b}}} \left(\bar{\Bell}_1,\bar{\Bell}_2,\bar{\Bell}_3 +\Bell^{''} \right)
    \Delta^{(3)}_{\ell_1^{'\mathrm{b}}, \ell_2^{'\mathrm{b}}, \ell_3^{'\mathrm{b}}} \left(\bar{\Bell}'_1,\bar{\Bell}'_2,\bar{\Bell}'_3 - \Bell^{''} \right)}{\Omega_{\text{sky}}^{2}\ N_{\text{tri.}}\left(\ell_1^{\mathrm{b}}, \ell_2^{\mathrm{b}}, \ell_3^{\mathrm{b}}\right)N_{\text{tri.}}\left(\ell^{'\mathrm{b}}_1,\ell^{'\mathrm{b}}_2,\ell^{'\mathrm{b}}_3\right)}. \label{NeedThis2}
\end{multline}
\begin{equation}
\text{Cov}\Big[\hat{P}^{\mathrm{W}}_{ij}\big(\ell^{\mathrm{b}}\big),\hat{B}^{\mathrm{W}}_{i'j'k'}(\ell^{\mathrm{b}}_1,\ell^{\mathrm{b}}_2,\ell^{\mathrm{b}}_3) \Big]_{\text{NG,P5}} =
     \sum_{\Bell'}\sum_{\Bell'_1,\Bell'_2,\Bell'_3}\int\frac{d^2\Bell^{''}}{\left( 2\pi\right)^2}|W(\Bell^{''})|^2
    P_{iji'j'k'}\left(\bar{\Bell}_1,\bar{\Bell}_2,\bar{\Bell}_3 + \Bell^{''}
    ,\bar{\Bell}',-\bar{\Bell}'- \Bell^{''}\right)
    \frac{\Delta^{(2)}_{\ell^{\mathrm{b}}}\left(\Bell\right)\Delta^{(3)}_{\ell_1^{\mathrm{b}}, \ell_2^{\mathrm{b}}, \ell_3^{\mathrm{b}}} \left(\Bell_1,\Bell_2,\Bell_3 \right)}{\Omega^2_{\text{sky}}\  N\left(\ell^{\mathrm{b}}\right)N_{\text{tri.}}\left(\ell_1^{\mathrm{b}}, \ell_2^{\mathrm{b}}, \ell_3^{\mathrm{b}}\right)}. \label{NeedThis3}
\end{equation}
Similarly to the power spectrum case, the Limber approximation allows to determine the convergence polyspectra via a line-of-sight integration of the matter density ones, at every redshift. The consistency relations allow the following decomposition \citep{2018PhRvD..97d3532C}
\begin{align}
\label{NGsscBB}
    P\left(\vec{k}_1,\vec{k}_2,\vec{k}_3 + \vec{p}
    ,\vec{k}'_1,\vec{k}'_2,\vec{k}'_3 - \vec{p}\right) &\approx  P\left(\vec{k}_1,\vec{k}_2,\vec{k}_3 
    ,\vec{k}'_1,\vec{k}'_2,\vec{k}'_3 \right) + \frac{\partial B(\vec{k}_1,\vec{k}_2,\vec{k}_3|\delta_{\mathrm{b}})}{\partial\delta_{\mathrm{b}}}\frac{\partial B(\vec{k}'_1,\vec{k}'_2,\vec{k}'_3|\delta_{\mathrm{b}})}{\partial\delta_{\mathrm{b}}}\ P^{\text{lin}}\left(p\right),\\
\label{NGsscBP}    P\left(\vec{k}_1,\vec{k}_2,\vec{k}_3 + \vec{p}
    ,\vec{k}',-\vec{k}'-\vec{p}\right) &\approx  P\left(\vec{k}_1,\vec{k}_2,\vec{k}_3,\vec{k}',-\vec{k}'\right) + \frac{\partial B(\vec{k}_1,\vec{k}_2,\vec{k}_3|\delta_{\mathrm{b}})}{\partial\delta_{\mathrm{b}}}\frac{\partial P(\vec{k}'|\delta_{\mathrm{b}})}{\partial\delta_{\mathrm{b}}}\ P^{\text{lin}}\left(p\right).
\end{align}
Therefore, we can complement Eq.~\eqref{CovBB} and Eq.~\eqref{Covpb} with the standard intra-survey covariance derived from the first terms
\begin{align}
   \text{Cov}\Big[\hat{B}^{\mathrm{W}}_{ijk}\big(\ell_1,\ell_2,\ell_3\big),\hat{B}^{\mathrm{W}}_{i'j'k'}\left(\ell'_1,\ell'_2,\ell'_3\right) \Big]_{\text{NGins,6P}}   &\approx  \frac{1}{\Omega_{\mathrm{sky}}}P_{ijki'j'k'}\left(\ell_1,\ell_2,\ell_3,\ell'_1,\ell'_2,\ell'_3\right), \label{Ins6P}\\
   \text{Cov}\Big[P_{ij}\left(\ell\right),B_{i'j'k'}&\left(\ell_1,\ell_2,\ell_3\right) \Big]_{\text{NGins,5P}} \approx \frac{1}{\Omega_{\mathrm{sky}}}P_{iji'j'k'} \left(\ell,-\ell,\ell_1,\ell_2,\ell_3\right). \label{Ins5P}
\end{align}
The super-survey ones instead come from the projected responses and we can nicely introduce a much more general expression for them
\begin{multline}
\label{WLSSCflatskyLimber}
\text{Cov}\left[\hat{P}^{\mathrm{W}}_{i_1,\dots,i_n} \left(\Bell_1, \dots,\Bell_n\right),\hat{P}^{\mathrm{W}}_{i'_1,\dots,i'_{n'}} \left(\Bell'_1, \dots,\Bell'_{n'}\right)\right]_{\text{NGssc}} = \\
=\int _0^{\infty} \mathrm{d}\chi\ \mathcal{T}\left(i_1,\dots,i_n;\chi\right) \ \mathcal{T}\left(i'_1,\dots,i'_{n'};\chi\right)  \frac{\partial  P \big(\mathbf{k}\big(\Bell_1,\chi\big), \dots,\mathbf{k}\big(\Bell_n,\chi\big)  |\delta_{\mathrm{b}}\big)}{\partial \delta_{\mathrm{b}}} \frac{\partial  P \big(\mathbf{k}\big(\Bell'_1,\chi\big), \dots,\mathbf{k}\big(\Bell'_{n'},\chi\big) |\delta_{\mathrm{b}}\big)}{\partial \delta_{\mathrm{b}}} \sigma^2_{W}\left(\chi\right)
\end{multline}
where the projection functions $\mathcal{T}$ are defined in Eq.~\eqref{nProjection}. Also in Eqs.~\eqref{Ins6P} and ~\eqref{Ins5P} , we assumed the 6- and the 5-point correlation functions to be well approximated by the 1-halo term. Therefore, we kept their dependence on just the magnitudes of the angular multipoles involved. The expression for the super-sample term \eqref{WLSSCflatskyLimber} is consistent with the the more general result in \cite{2016JCAP...08..005L}.

To estimate the variance $\sigma^2_{W}$ of the long mode at every redshift, we use a cylindrical mask in real space for the single observed patch, the Fourier expansion of which can be derived as done in \cite{2007PhRvD..76l3013L}
\begin{equation}
\label{Mask}
\text{W}_{\text{cyl}}\left(\mathbf{k},\delta_{\chi},\hat{\chi}\right) = 2 \exp\left(i\ k_{\parallel}\hat{\chi}\right) j_0\left(\frac{1}{2}k_{\parallel}\delta_{\chi}\right)\frac{\text{J}_1\big(k_{\perp}\hat{\chi}\ \Theta_{\text{sky}}\big)}{k_{\perp}\hat{\chi}\ \Theta_{\text{sky}}}. 
\end{equation}
In particular Eq.~\eqref{Mask} is the Fourier transform of the selection function for a comoving cylindrical volume of depth $\delta_{\chi}$, centered in $\hat{\chi}$ and derived under the assumption of a slowly varying $\chi$ and Hubble factor $H\left(\chi\right)$ within $\delta_{\chi}$. Compared to the expression in \cite{2007PhRvD..76l3013L} we are omitting the photometric error related component.  The special functions $j_0$ and $\text{J}_1$ are respectively the $0^{\text{th}}$ order spherical Bessel function and the $1^{\text{st}}$ order Bessel function of the first kind.
Since we are interested in computing the variance of the matter field at a specific redshift, we compute the variance $\sigma_W^2(\chi)$ over a disk-like volume in the limit $\delta_{\chi}\to 0,\ \hat{\chi}\to\chi$ of Eq.~\eqref{Mask}
\begin{equation}
    \sigma_W^2\left(\chi\right) = 4\ \int \frac{{\mathrm{d}}^2\boldsymbol{\epsilon}_{\perp}}{\left(2\pi\right)^2} \left[\frac{\text{J}_1\big(\boldsymbol{\epsilon}_{\perp}\chi\ \Theta_{\text{sky}}\big)}{\boldsymbol{\epsilon}_{\perp}\chi\ \Theta_{\text{sky}}} \right]^2\ P^{\text{lin.}}\left(\boldsymbol{\epsilon}_{\perp},\chi\right). 
\end{equation}


\bsp	
\label{lastpage}
\end{document}